\documentclass[10pt,twocolumn]{IEEEtran}

\usepackage[mathscr]{eucal}
\usepackage[cmex10]{amsmath}
\usepackage{epsfig,epsf,psfrag}
\usepackage{amssymb,amsmath,amsthm,amsfonts,latexsym}
\usepackage{amsmath,graphicx,bm,xcolor,url}
\usepackage[caption=false]{subfig} 
\usepackage{fixltx2e}
\usepackage{array}
\usepackage{verbatim}
\usepackage{bm}
\usepackage{algorithmic, cite}
\usepackage{algorithm}
\usepackage{verbatim}
\usepackage{textcomp}
\usepackage{mathrsfs}
\usepackage{epstopdf}

\catcode`~=11 \def\UrlSpecials{\do\~{\kern -.15em\lower .7ex\hbox{~}\kern .04em}} \catcode`~=13 

\allowdisplaybreaks[1]
 
\newcommand{\nn}{\nonumber}

\newcommand{\calA}{\mathcal{A}}
\newcommand{\calB}{\mathcal{B}}
\newcommand{\calC}{\mathcal{C}}

\newcommand{\calN}{\mathcal{N}}

\newcommand{\calP}{\mathcal{P}}
\newcommand{\calQ}{\mathcal{Q}}

\newcommand{\calT}{\mathcal{T}}

\newcommand{\bI}{\mathbf{I}}

\newcommand{\bJ}{\mathbf{J}}

\newcommand{\bu}{\mathbf{u}}
\newcommand{\bU}{\mathbf{U}}

\newcommand{\bV}{\mathbf{V}}

\newcommand{\bx}{\mathbf{x}}

\newcommand{\by}{\mathbf{y}}

\newcommand{\bz}{\mathbf{z}}

\newcommand{\rmd}{\mathrm{d}}

\newcommand{\rme}{\mathrm{e}}

\newcommand{\rmf}{\mathrm{f}}

\newcommand{\rmP}{\mathrm{P}}

\newcommand{\rmQ}{\mathrm{Q}}

\newcommand{\bbE}{\mathbb{E}}

\newcommand{\bbN}{\mathbb{N}}

\newcommand{\bbR}{\mathbb{R}}

\DeclareMathAlphabet{\mathbsf}{OT1}{cmss}{bx}{n}
\DeclareMathAlphabet{\mathssf}{OT1}{cmss}{m}{sl}

\newcommand{\rvC}{\mathsf{C}}

\newcommand{\rvV}{\mathsf{V}}

\DeclareSymbolFont{bsfletters}{OT1}{cmss}{bx}{n}  
\DeclareSymbolFont{ssfletters}{OT1}{cmss}{m}{n}
\DeclareMathSymbol{\bsfGamma}{0}{bsfletters}{'000}
\DeclareMathSymbol{\ssfGamma}{0}{ssfletters}{'000}
\DeclareMathSymbol{\bsfDelta}{0}{bsfletters}{'001}
\DeclareMathSymbol{\ssfDelta}{0}{ssfletters}{'001}
\DeclareMathSymbol{\bsfTheta}{0}{bsfletters}{'002}
\DeclareMathSymbol{\ssfTheta}{0}{ssfletters}{'002}
\DeclareMathSymbol{\bsfLambda}{0}{bsfletters}{'003}
\DeclareMathSymbol{\ssfLambda}{0}{ssfletters}{'003}
\DeclareMathSymbol{\bsfXi}{0}{bsfletters}{'004}
\DeclareMathSymbol{\ssfXi}{0}{ssfletters}{'004}
\DeclareMathSymbol{\bsfPi}{0}{bsfletters}{'005}
\DeclareMathSymbol{\ssfPi}{0}{ssfletters}{'005}
\DeclareMathSymbol{\bsfSigma}{0}{bsfletters}{'006}
\DeclareMathSymbol{\ssfSigma}{0}{ssfletters}{'006}
\DeclareMathSymbol{\bsfUpsilon}{0}{bsfletters}{'007}
\DeclareMathSymbol{\ssfUpsilon}{0}{ssfletters}{'007}
\DeclareMathSymbol{\bsfPhi}{0}{bsfletters}{'010}
\DeclareMathSymbol{\ssfPhi}{0}{ssfletters}{'010}
\DeclareMathSymbol{\bsfPsi}{0}{bsfletters}{'011}
\DeclareMathSymbol{\ssfPsi}{0}{ssfletters}{'011}
\DeclareMathSymbol{\bsfOmega}{0}{bsfletters}{'012}
\DeclareMathSymbol{\ssfOmega}{0}{ssfletters}{'012}

\newcommand{\hatp}{\hat{p}}
\newcommand{\hatP}{\hat{P}}

\newcommand{\tilP}{\tilde{P}}

\newcommand{\hatW}{\hat{W}}

\newcommand{\tilX}{\tilde{X}}

\newcommand{\tilZ}{\widetilde{Z}}

\newcommand{\barp}{\bar{p}}

\newcommand{\barP}{\bar{P}}

\newcommand{\barX}{\bar{X}}

\newcommand{\barZ}{\bar{Z}}

\newcommand{\eps}{\varepsilon}

\def\fndot{\, \cdot \,}

\newcommand{\iid}{i.i.d.\ }

\DeclareMathOperator*{\argmin}{arg\,min}

\DeclareMathOperator{\var}{\mathsf{Var}}

\DeclareMathOperator{\cov}{\mathsf{Cov}}

\newtheorem{theorem}{Theorem}

\newcommand{\qednew}{\nobreak \ifvmode \relax \else
      \ifdim\lastskip<1.5em \hskip-\lastskip
      \hskip1.5em plus0em minus0.5em \fi \nobreak
      \vrule height0.75em width0.5em depth0.25em\fi}

\usepackage{epsfig} 
\usepackage{epstopdf}  
\usepackage{overpic}

\usepackage{dsfont}
\usepackage{xspace}

\usepackage[ colorlinks = true,
             linkcolor = blue,
             urlcolor  = blue,
             citecolor = red,
             anchorcolor = green,
]{hyperref}

\newcommand{\red}{\textcolor[rgb]{0,0,0}}
\newcommand{\blue}{\textcolor[rgb]{0,0,0}}

\newcommand{\purple}{\textcolor[rgb]{0.0,0,0.0}}

 \usepackage{cite}

\let\iid\undefined
\newcommand{\iid}{i.i.d.\@\xspace}

\let\tilP\undefined
\newcommand{\tilP}{\widetilde{P}}
\newcommand{\lefto}{\mathopen{}\left}

\begin{document} 
\flushbottom
\title{The Dispersion of Nearest-Neighbor Decoding for Additive Non-Gaussian Channels} 

\author{Jonathan Scarlett,~\IEEEmembership{Member,~IEEE}, $\,\,\,$ 
        Vincent Y.~F.~Tan,~\IEEEmembership{Senior Member,~IEEE}, \\Giuseppe Durisi,~\IEEEmembership{Senior Member,~IEEE} 
\thanks{J.~Scarlett  is   with the Laboratory for Information and Inference Systems, \'Ecole Polytechnique F\'ed\'erale de Lausanne, CH-1015, Switzerland
(email:\,jmscarlett@gmail.com).}
\thanks{V.~Y.~F.~Tan is with the   Department of Electrical and Computer Engineering and the Department of Mathematics, National University of Singapore
(email:\,vtan@nus.edu.sg).}
\thanks{G.~Durisi is with the Department of Signals and Systems, Chalmers University of Technology, Sweden (email:\,durisi@chalmers.se).  }
\thanks{The work of V.~Y.~F.~Tan is supported in part by a Singapore Ministry of Education (MOE) Tier 2 grant (R-263-000-B61-113). 
The work of G. Durisi was supported in part by the Swedish Research Council (VR) under grant no. 2012-4571.
} \thanks{This paper was presented in part at the 2016 International Zurich Seminar on Communications and the 2016 International Symposium on Information Theory, Barcelona, Spain.}
}
 
\IEEEpeerreviewmaketitle
 
\maketitle

\begin{abstract}  
We study the second-order asymptotics of information transmission using random Gaussian codebooks and nearest neighbor (NN) decoding over a power-limited stationary memoryless additive non-Gaussian noise channel. We show  that the dispersion term depends on the non-Gaussian noise only through its second and fourth moments, \blue{thus complementing the capacity result (Lapidoth, 1996), which depends only on the second moment.}
Furthermore, we characterize the second-order asymptotics of point-to-point codes over \red{$K$-sender interference networks with non-Gaussian additive noise}. 
Specifically, we assume that each user's codebook is Gaussian and that NN decoding is employed, i.e., that interference from the \red{$K-1$ unintended users (Gaussian interfering signals)}  is treated as noise at each decoder.
We show that while the first-order term in the asymptotic expansion of the maximum number of messages depends on the power \red{of} the interferring codewords only through their sum, this does not hold for the second-order term.
\end{abstract} 

 
\section{Introduction}
In second-order asymptotic analyses \red{for channel coding}, a line of work pioneered by Strassen~\cite{Strassen} and extended by Hayashi~\cite{Hayashi09} and Polyanskiy, Poor and Verd\'u~\cite{PPV10} among others, one seeks a two-term approximation of the maximum possible number of messages $M^*(n,\eps)$ that can be transmitted with an average probability of error no larger  than $\eps$ using a stationary memoryless channel $n$ times, \red{as $n$ tends to infinity}. 
This analysis is useful because it serves as a good approximation to the finite-blocklength fundamental limits of information transmission over various channels for error probabilities of practical interest~\cite{PPV10}.
For a real AWGN channel with unit-variance noise, and under the assumption that the $n$-dimensional codewords have a power not exceeding $nP$, it is well-known that $M^*(n,\eps)$ can be approximated as~\cite{Hayashi09,PPV10}
\begin{equation}\label{eq:second_order_expansion_gaussian}
\log M^*(n,\eps) = n\rvC(P) - \sqrt{n\rvV(P)}\rmQ^{-1}(\eps) + O(\log{n}),
\end{equation}
where  $\rmQ(a) := ({1}/{\sqrt{2\pi}})\int_{a}^{\infty}e^{-t^2/2}\, \rmd t$ is the complementary cumulative distribution function (CDF) of a standard Gaussian,  $\rmQ^{-1}(\cdot)$ is its inverse,
%
$\rvC(P) = ({1}/{2})\log(P+1)$ [nats per channel use]
%
is the channel capacity, and
\begin{IEEEeqnarray}{rCL}\label{eq:gaussian_dispersion}
  \rvV(P)=\frac{P(P+2)}{2(P+1)^2}\quad\mbox{[nats$^2$  per channel use]}
\end{IEEEeqnarray}
is the channel dispersion.  \blue{An expression of the same form as~\eqref{eq:gaussian_dispersion}  was also derived by Shannon~\cite{Sha59b} in his study of the optimal asymptotic error probability of transmission over an AWGN channel at rates close to capacity.}

\emph{Spherical Gaussian codebooks} (\blue{henceforth also referred to as {\em shell codebooks} or simply {\em shell codes}}), where each codeword is  uniformly distributed  on a sphere of radius $\sqrt{nP}$, achieve~\eqref{eq:second_order_expansion_gaussian}.  
Furthermore, the \emph{nearest-neighbor} (NN) or \emph{minimum distance} decoding rule is optimal in the sense that this rule minimizes the error probability.  
In this paper, we also study the performance of {\em \iid Gaussian codebooks}, which achieve $\rvC(P)$ as well, but are not dispersion-optimal (i.e., they do not achieve the second term in \eqref{eq:second_order_expansion_gaussian}). 
Nevertheless, they provide a basis for comparison to other coding schemes. 

A natural question then beckons. 
What is the maximum rate we can achieve if the codebook is constrained to be Gaussian and the decoder uses the NN rule, but the noise is non-Gaussian? See Fig.~\ref{fig:non_gauss}.  This question is relevant in situations when one knows how to combat Gaussian noise, and seeks to adopt the same design for the non-Gaussian case, despite the inherent mismatch.  

Lapidoth~\cite{Lapidoth} provided a ``first-order-asymptotics'' answer to this question by showing that the Gaussian capacity $\rvC(P)$ is also the largest rate achievable over unit-variance non-Gaussian stationary ergodic additive channels when i.i.d.~or spherical Gaussian codebooks and NN decoding are used. Lapidoth also showed that the rate of communication cannot exceed $\rvC(P)$ \red{using this encoding-decoding structure} even if one allows for a non-vanishing error probability $\eps\in (0,1)$, i.e., Lapidoth proved a {\em strong converse} with respect to the random codebook. 
In some sense, these results say  that Gaussian codebooks and NN  decoding, although possibly suboptimal for the case of non-Gaussian noise, form a robust communication scheme---\blue{the target rate $\rvC(P)$ is achieved even when the assumption of Gaussian noise fails to hold.}

\subsection{Main Contributions and Proof Technique}
In this paper, we extend Lapidoth's result in the direction of second-order asymptotics
by determining the analogue of $\rvV(P)$ in~\eqref{eq:second_order_expansion_gaussian} for additive non-Gaussian noise channel, when we use a Gaussian codebooks and a  NN decoder.
Specifically, we show that when the non-Gaussian noise is \iid and has a finite fourth moment $\xi$ and a finite sixth moment, the dispersion for the case of spherical Gaussian codebooks and NN decoder is given by
\begin{equation}
\rvV(P,\xi):=\frac{P^2 (\xi-1)+4P}{4(P+1)^2}. \label{eqn:disp_intro}
\end{equation}
This means that the rate of convergence to $\rvC(P)$ depends on the fourth moment of the noise distribution, or more generally, on the \emph{kurtosis}, i.e., the ratio between the fourth-moment and the square of the second moment.  The higher the kurtosis, the slower the convergence. We also perform a similar analysis for the case that the codebook is \iid Gaussian, and obtain a expression for the dispersion that is, in general, larger (worse) than that in \eqref{eqn:disp_intro}. 

In addition, motivated by work by Baccelli, El Gamal and Tse~\cite{bacelli} and Bandemer, El Gamal and Kim~\cite{Bandemer} on  communication over interference networks with point-to-point codes, we establish the  dispersion for the scenario where additional ``noise'' arises from unintended users equipped with Gaussian codebooks, and the NN decoding rule is used at the intended receiver. This means that interference is treated as noise.
For this case, whereas the first term in the second-order expansion of $\log M^*(n,\eps)$ is simply $n$ times the Gaussian channel capacity $\rvC(\cdot)$ with $P$ replaced by the signal-to-interference-and-noise ratio (SINR), the expression for the channel dispersion is more involved (see~\eqref{eqn:shell_disp2}).
In particular, it depends on the individual power of each interferer and not only on the total interference power. 
This investigation sheds lights on a setup that is appealing from a practical point of view, because the communication scheme that is analyzed has low complexity---each transmitter is constrained to use a Gaussian codebook and only simple point-to-point codes are considered.
In particular, more complex schemes such as  superposition~\cite{cover72} or Han-Kobayashi~\cite{Han81} coding are not permitted. 

One of the main tools in our second-order analysis is the \emph{Berry-Esseen theorem for functions of random vectors} (e.g.,~see \cite[Prop.~1]{Iri15}). To ensure the paper is self-contained, we have restated this as Theorem \ref{thm:berry} in Appendix \ref{app:be_func}.  
This result, which is also known as the {\em multivariate delta method}~\cite{cra99} in statistics,   has previously been used to obtain an inner bound  to the second-order rate region for Gaussian multiple access channel by MolavianJazi and Laneman~\cite{Mol13c}. 
Moreover, it has also been successfully employed in  the analyses of the second-order asymptotics of certain classes of Gaussian  interference channels~\cite{Quoc15}, as well as the third-order asymptotics  for fixed-to-variable length universal compression of \iid~\cite{KosutSankar1} and  Markov~\cite{Iri15} sources. 
\subsection{Paper Outline}
The remainder of the paper is structured as follows. 
In  Section~\ref{sec:p2p}, we present our second-order asymptotic results for point-to-point additive non-Gaussian noise channels, for the setup where the encoder outputs a codeword from a random Gaussian codebook (to be made precise therein) and the decoder is constrained to return the codeword that is closest in Euclidean distance to the channel output. We consider both shell and \iid codes. In Section~\ref{sec:int_network}, we study interference networks and derive similar second-order asymptotic expansions. 
The proofs of our main results are presented in Sections~\ref{sec:prf_add} and \ref{sec:prf_add2}. We conclude the paper in Section~\ref{sec:conclu}. 
 \begin{figure}
 \centering
 \includegraphics[width = 1.0\columnwidth]{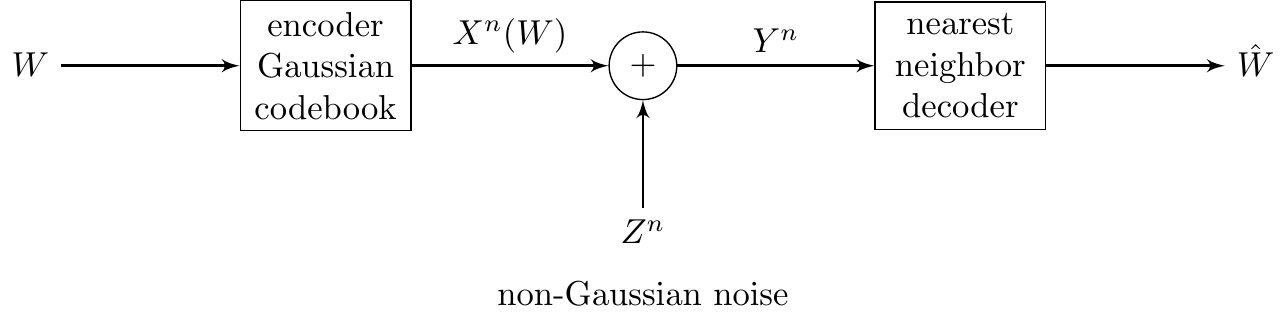}
 \caption{Additive non-Gaussian noise channel.}
 \label{fig:non_gauss}
 \end{figure}

\section{Point-To-Point Channels }  \label{sec:p2p}
\subsection{System Model and Definitions}
Consider the  point-to-point additive-noise channel
\begin{equation}\label{eq:channel}
Y^n = X^n + Z^n,
\end{equation}
where $X^n$ is the input vector and $Z^n$ is the noise vector over $n$ scalar channel uses. See Fig.~\ref{fig:non_gauss}.
Throughout, we shall focus exclusively on Gaussian codebooks. 
More precisely, we consider {\em shell codes} for which $X^n$ is uniformly distributed on a sphere with radius $\sqrt{nP}$, i.e.,
\begin{align}
X^n \sim \rmf_{X^n}^{(\mathrm{shell})}(\bx) := \frac{\delta ( \|\bx\|^2 - nP) }{S_n(\sqrt{nP}) }. \label{eqn:shell}
\end{align}
Here, $\delta(\cdot)$ is the Dirac delta function, and $S_n(r) = 2\pi^{n/2}r^{n-1}/\Gamma(n/2 )$ is the surface area of a radius-$r$ sphere in $\bbR^n$ where $\Gamma(\cdot)$ denotes the Gamma function. 
This random coding distribution is second- and third-order optimal~\cite{TanTom13a}. 
For the purpose of comparison, we shall also consider {\em \iid Gaussian codes}, in which each component of $X^n$ is distributed according to a zero-mean, variance $P$ normal distribution, i.e., 
\begin{equation}
X^n \sim \rmf_{X^n}^{({\mathrm{iid}})}(\bx) := \prod_{i=1}^n \frac{1}{\sqrt{2\pi P}}\exp\Bigl( -\frac{x_i^2}{2 P} \Bigr). \label{eqn:iid}
\end{equation}
This random coding distribution achieves $\rvC(P)$ but not $\rvV(P)$. \blue{Note that while the quantity $P$ in~\eqref{eqn:shell} can be interpreted as a \emph{peak} or \emph{per-codeword} power constraint (e.g., see \cite[Eq.~(192)]{PPV10}), this does not hold for \eqref{eqn:iid}.  There, $P$ should instead be interpreted as the power averaged over the random codebook.}

The noise $Z^n$ is generated according to a stationary and memoryless process that does not depend on the channel input:
\begin{equation}
Z^n\sim \rmP_{Z^n}(\bz) =\prod_{i=1}^n \rmP_Z(z_i)
\end{equation}
for some $\rmP_Z$. Hence, the $n^{\mathrm{ th}}$ extension of the channel is
\begin{equation}
\rmP_{Y^n|X^n}(\by|\bx)=\prod_{i=1}^n\rmP_{Y|X}(y_i|x_i)=\prod_{i=1}^n \rmP_Z(y_i-x_i).
\end{equation}
The distribution $\rmP_Z$ does not need to be Gaussian \blue{nor even zero-mean}; the only assumptions are the following:
\begin{equation}
  \bbE\bigl[ Z^2\bigr] = 1,  \quad \xi:=\bbE\bigl[Z^4\bigr]<\infty,\quad\bbE\bigl[Z^6\bigr]<\infty.  \label{eqn:moments}
\end{equation}
The assumption that the  second moment is equal to $1$ \blue{is made for normalization purposes}.  As we shall see, the  assumption that the fourth moment is finite is crucial.
 \purple{Note that the finiteness of the fourth moment implies by the Cauchy-Schwarz inequality that the second moment is finite as well.} 
 The assumption that the sixth moment is finite is made only for technical reasons (see Appendix \ref{app:moderate}). 

Given either a shell or an \iid codebook consisting of $M \in\bbN$ random codewords $\calC:=\{X^n(1),\ldots, X^n(M)\}$, we consider an NN (or minimum distance) decoder that outputs the message $\hatW $ whose corresponding codeword is closest in Euclidean distance to the channel output $Y^n$, i.e.,
\begin{equation}
\hatW := \argmin_{ w \in [1:M ]}\,\,  \| Y^n-X^n(w) \|.\label{eqn:NN}
\end{equation}
\red{This  decoder  is optimal if the noise is Gaussian~\cite{Lapidoth, Cov06}}, but may not be so in the more general setup considered here. 

We define the {\em average probability of error}  as 
\begin{equation}
\barp_{\rme, n}:= \Pr[\hatW\ne W].
\end{equation}
This probability is averaged over the uniformly distributed message $W$, the random  codebook $\calC$, and the channel noise $Z^n$. 
Note that in traditional channel-coding analyses~\cite{Hayashi09,PPV10}, the probability of error is  averaged only over $W$ and $Z^n$.
Similar to~\cite{Lapidoth}, the additional averaging over the codebook $\calC$ is required here to establish ensemble-tightness results for the two classes of Gaussian codebooks considered in this paper.

Let $M^*_{\mathrm{shell}}(n,\eps,P;\rmP_Z)$ be the maximum number of messages that can be transmitted using  a shell codebook over the channel~\eqref{eq:channel}  with average error probability $\barp_{\rme, n}$ no larger than $\eps\in (0,1)$, when the noise is distributed according to $\rmP_Z$. 
Let $M^*_{\mathrm{iid}}(n,\eps,P;\rmP_Z)$ be the analogous quantity for the case of \iid Gaussian codebooks. 
Lapidoth~\cite{Lapidoth} showed that  for all $\eps\in (0,1)$ and $\dagger\in\{\mathrm{shell},\mathrm{iid}\}$, 
\begin{equation}
\lim_{n\to\infty}\frac{1}{n}\log M^*_{\dagger}(n,\eps,P;\rmP_Z)=\rvC(P) \label{eqn:lap}
\end{equation}
regardless of the  distribution of $\rmP_Z$. 
Note that Lapidoth's result~\cite[Th.~1]{Lapidoth} holds \blue{under more general noise models than that of the present paper}. 
\purple{In particular, it allows for distributions with memory}.

\subsection{Main Result}
 In the following theorem, we provide a second-order asymptotic expansion of $\log M^*_\dagger(n,\eps,P;\rmP_Z)$. 
\begin{theorem} \label{thm:disp}
Consider a noise distribution $\rmP_Z$ with statistics as in \eqref{eqn:moments}. For shell codes, 
\begin{align}
&\log M^*_{\mathrm{shell}}(n,\eps,P;\rmP_Z) \nn\\*
&\quad =n\rvC(P) - \sqrt{n \rvV_{\mathrm{shell}}(P,\xi)}\rmQ^{-1}(\eps) + O(\log n), \label{eqn:asymp_shell}
\end{align}
where the shell dispersion is 
\begin{equation}
\rvV_{\mathrm{shell}}(P,\xi):=\frac{P^2 (\xi-1)+4P}{4(P+1)^2}. \label{eqn:disp_shell}
\end{equation}
Moreover, for \iid codes, 
\begin{align}
&\log M^*_{\mathrm{iid}}(n,\eps,P;\rmP_Z) \nn\\*
&\quad =n\rvC(P) - \sqrt{n\rvV_{\mathrm{iid}}(P,\xi)}\rmQ^{-1}(\eps) +O(\log n), \label{eqn:asymp_iid}
\end{align}
where the \iid dispersion is 
\begin{equation}
\rvV_{\mathrm{iid}}(P,\xi):=\frac{P^2(\xi+1)+4P}{4(P+1)^2}.\label{eqn:disp_iid}
\end{equation}
\end{theorem}

The proof of Theorem \ref{thm:disp} is given in Section \ref{sec:prf_add}. 

\subsection{Remarks on Theorem \ref{thm:disp}}

The second-order terms in the asymptotic expansions of $\log M^*_{\mathrm{shell}}(n,\eps,P;\rmP_Z)$ and $\log M^*_{\mathrm{iid}}(n,\eps,P;\rmP_Z)$ only depend on the distribution $\rmP_Z$ through its second and fourth moments. 
If $Z$ is Gaussian, then the fourth moment $\xi$ is equal to $3$ and we recover from~\eqref{eqn:disp_shell} the  Gaussian dispersion~\eqref{eq:gaussian_dispersion}.

Comparing~\eqref{eqn:disp_shell} with~\eqref{eq:gaussian_dispersion}, we see that noise distributions $\rmP_Z$ with higher fourth moments than Gaussian (e.g., Laplace) result in a slower convergence to $\rvC(P)$ \blue{for $\epsilon < \frac{1}{2}$}. 
Conversely, distributions with smaller fourth moment than Gaussian (e.g., Bernoulli) result in a faster convergence to $\rvC(P)$.

For the case of \iid Gaussian codes  (i.e., $X^n\sim  \rmf_{X^n}^{({\mathrm{iid}})}$), when $Z$ is Gaussian,  \textcolor{black}{$\xi = 3$, and so from \eqref{eqn:disp_iid}, we obtain}
\begin{equation}
\rvV_{\mathrm{iid}}(P) = \frac{P}{P+1}. \label{eqn:rice} 
\end{equation}
An expression of the same form as~\eqref{eqn:rice} was derived by Rice~\cite{Rice}, who used \iid Gaussian codes to establish a lower bound on  the error exponent  (reliability function) for AWGN channels at rates close to capacity.

Note finally that 
\begin{equation}
\rvV_{\mathrm{iid}}(P,\xi)=\rvV_{\mathrm{shell}}(P,\xi) +\frac{1}{2}\Big(\frac{ P }{  P+1 }\Big)^2.
\end{equation}
Since this implies that $\rvV_{\mathrm{shell}}(P,\xi) \le\rvV_{\mathrm{iid}}(P,\xi)$, we conclude that shell codes are superior to \iid codes \blue{ for $\epsilon < {1}/{2}$}, because they yield a smaller dispersion.  \blue{Note that $\log M^*_{\mathrm{iid}}$ in \eqref{eqn:asymp_iid} is only achievable for $\eps > {1}/{2}$} \purple{under the assumption that $P$ is the power averaged over the random codebook; for the usual setting with a per-codeword power constraint,~\eqref{eqn:asymp_iid} is not achievable, as this would violate known converse bounds \cite{Hayashi09,PPV10,TanTom13a} for $\eps>1/2$.}  

\section{Interference Networks}\label{sec:int_network} 
\subsection{System Model  and Definitions}
We assume that $K$ sender-receiver pairs operate concurrently over the same additive noise channel.
Similarly to Section~\ref{sec:p2p}, the additive noise $Z^n$ is \iid but possibly non-Gaussian.
The senders use Gaussian codebooks with powers $\{P_j\}_{j=1}^K$  (as in Section~\ref{sec:p2p}, we shall consider both shell and \iid codes) and all receivers use NN decoding. 
Hence, they treat the codewords from the unintended senders as additional noise. Note that for the case of shell codes, the resulting total additive noise (including interference) is no longer \iid, so Theorem~\ref{thm:disp} cannot be applied.
Although the communication strategy  described above may be suboptimal rate-wise compared to  more sophisticated strategies such as superposition~\cite{cover72} or Han-Kobayashi coding~\cite{Han81}, it is easy to implement since it relies exclusively on point-to-point channel codes~\cite{bacelli,Bandemer}.

 \begin{figure*}
\centering
\includegraphics[width = 1.35\columnwidth]{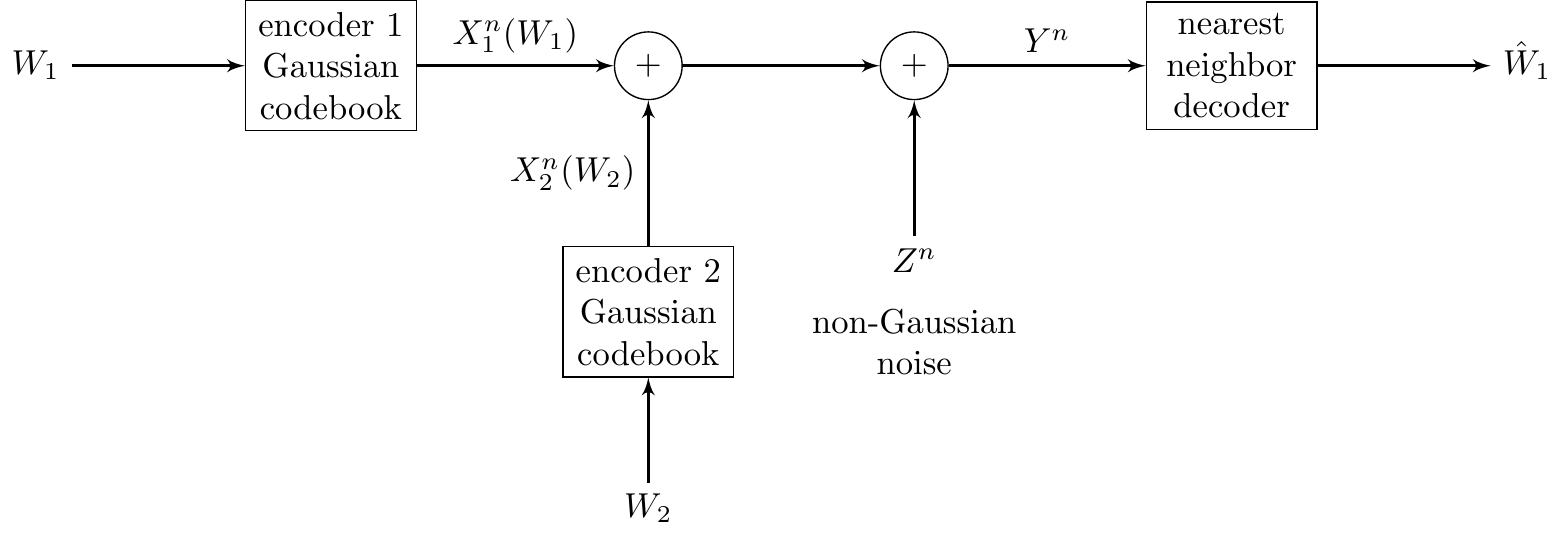}
\caption{Non-Gaussian noise plus interference.}
\label{fig:non_gauss_int}
\end{figure*}

Without loss of generality, we shall focus only on the signal at receiver $1$, which we indicate by
\begin{equation}\label{eq:channel_int}
Y^n = X_1^n+X_2^n+\ldots + X_K^n + Z^n.
\end{equation}
Here, $X_j^n$ denotes the codeword transmitted by the $j^\mathrm{th}$ sender; $X_1^n$ is the codeword intended for receiver 1.  
See Fig.~\ref{fig:non_gauss_int} for an illustration of this setting for the case when the number of senders $K$ is set to $2$. 

We consider two different   models for the codewords. 
In the first, each codeword $X_j^n$ follows a {\em shell distribution} as in \eqref{eqn:shell} with power $P_j$. 
In the second, each codeword $X_j^n$ follows an {\em \iid Gaussian distribution} as in \eqref{eqn:iid} with   power $P_j$. We define $M^*_{\mathrm{shell}}(n,\eps,\{P_j\}_{j=1}^K;\rmP_Z)$ to be the maximum number of messages that sender $1$  can transmit using shell codes  with  error probability no larger than $\eps\in (0,1)$ over the channel~\eqref{eq:channel_int}, when the receiver uses NN decoding, and, hence, treats $X_2^n,\ldots, X_K^n$ as noise. 
Similarly, let  $M^*_{\mathrm{iid}}(n,\eps,\{P_j\}_{j=1}^K;\rmP_Z)$ be the analogous quantity for the case of \iid Gaussian codebooks.  Let the SINR of the channel from sender $1$ to receiver $1$ be
\begin{equation}
\bar{P} := \frac{P_1}{ 1+\tilP  },\label{eqn:sinr}
\end{equation}
where the total power of the interfering codewords is 
\begin{equation}
\tilP:=\sum_{j=2}^K P_j.\label{eqn:int_power}
\end{equation}

\subsection{Main Result}
In Theorem~\ref{thm:disp2} below, we provide a second-order asymptotic expansion of $\log M^*_{\dagger}(n,\eps,\{P_j\}_{j=1}^K;\rmP_Z)$.
\begin{theorem}
\label{thm:disp2}
Consider a  noise distribution $\rmP_Z$ with statistics as in \eqref{eqn:moments}. 
 For shell codes, 
\begin{align}
& \log M^*_{\mathrm{shell}}(n,\eps,\{P_j\}_{j=1}^K;\rmP_Z)  \nn\\*
&\quad =n\rvC(\bar{P}) - \sqrt{n \rvV_{\mathrm{shell}}'( \{P_j\}_{j=1}^K ,\xi)}\rmQ^{-1}(\eps) +O(\log n) \label{eqn:interf_shell}
\end{align}
where the shell dispersion is 
\begin{align}
 &\rvV_{\mathrm{shell}}'( \{P_j\}_{j=1}^K ,\xi)  \nn\\*
& := \frac{P_1^2 (\xi  \! -\!1 \! +\! 4\tilP) + 4P_1 ( \tilP\!+\! 1)^3 + 4P_1^2\sum_{2\le i<j\le K}P_iP_j}{4( \tilP\!+\!1)^2 (P_1\!+\!\tilP\!+\! 1)^2}. \label{eqn:shell_disp2}
\end{align}
Moreover, for \iid codes, 
\begin{align}
& \log M^*_{\mathrm{iid}}(n,\eps,\{P_j\}_{j=1}^K;\rmP_Z)  \nn\\*
 &\quad  =n\rvC(\bar P) - \sqrt{n \rvV_{\mathrm{iid}} (\bar P,\xi') }\rmQ^{-1}(\eps) + O(\log n)  \label{eqn:interf_iid}
\end{align}
where $\rvV_{\mathrm{iid}}(\cdot,\cdot)$ is defined in \eqref{eqn:disp_iid}, and
\begin{equation}
\xi':= \frac{3\tilP^2 + 6\tilP +\xi}{(\tilP+1)^2}. \label{eqn:xi_prime}
\end{equation}
\end{theorem} 
The  proof of this result can be found in Section \ref{sec:prf_add2}.

\subsection{Remarks on Theorem \ref{thm:disp2}}

\blue{As a sanity check, we first notice that Theorem \ref{thm:disp2} reduces to Theorem \ref{thm:disp} upon setting $K=1$, which immediately yields $\widetilde P = 0$, $\bar P = P$, and $\xi' = \xi$.}

As expected, the first-order term in the asymptotic expansion is $\rvC(\bar{P})$, where $\bar{P}$ is the SINR defined in~\eqref{eqn:sinr}. 
The second-order term for the \iid  case can be obtained in a straightforward manner from Theorem~\ref{thm:disp} (see Section~\ref{sec:prf_inter_iid}). 
This is because the effective noise, $X_2^n+\ldots+X_K^n + Z^n$, is  \iid Gaussian, but of variance  $\tilP+1$ instead of $1$. 
The second-order term for the shell case does not follow directly from Theorem~\ref{thm:disp}, because the total noise $X_2^n+\ldots+X_K^n + Z^n$, which is the sum of $K-1$ shell random vectors and a single \iid Gaussian random vector, is neither \iid nor shell distributed.

Observe that the cross term  in~\eqref{eqn:shell_disp2}, namely $\red{2}\sum_{2\le i<j\le K}P_iP_j = (\sum_{j=2}^K P_j)^2 - \sum_{j=2}^K P_j^2$ implies that the dispersion does not only depend on the sum $\tilP$ of the interferers' powers,  but also on the individual power $P_j$ of each interferer.
This phenomenon is not present in the \iid case. 
Since, by convexity, 
\begin{equation}
\purple{\frac{\tilP^2}{K-1} \leq \sum_{j=2}^{K}P_j^2\leq  \tilP^2} \label{eq:jensen_P}
 \end{equation} 
we conclude that the shell dispersion is maximized when all interferers have the same power, and \purple{minimized when only one interferer with power $\tilP$ is active.}
Through standard manipulation along with \eqref{eq:jensen_P}, one can also show that $\rvV_{\mathrm{shell}}'( \{P_j\}_{j=1}^K ,\xi) \leq \rvV_{\mathrm{iid}} (\bar P,\xi') $ for all $\{P_j\}_{j=1}^K$.

Finally, observe that if $Z$ is standard Gaussian, we have $\xi =\bbE  [Z^4]= 3$ and hence $\xi' = 3$ (see~\eqref{eqn:xi_prime}). 
Plugging this into the expression for $\rvV_{\mathrm{iid}}$ in~\eqref{eqn:disp_iid}, evaluated at $\barP$ and $\xi'$, we obtain as expected
\begin{equation}
\rvV_{\mathrm{iid}}(\barP,\xi') = \frac{\barP}{\barP+1}.
\end{equation}
This is in complete analogy to~\eqref{eqn:rice}. 
\begin{figure}
  \centering 
    \includegraphics[width=.99\columnwidth]{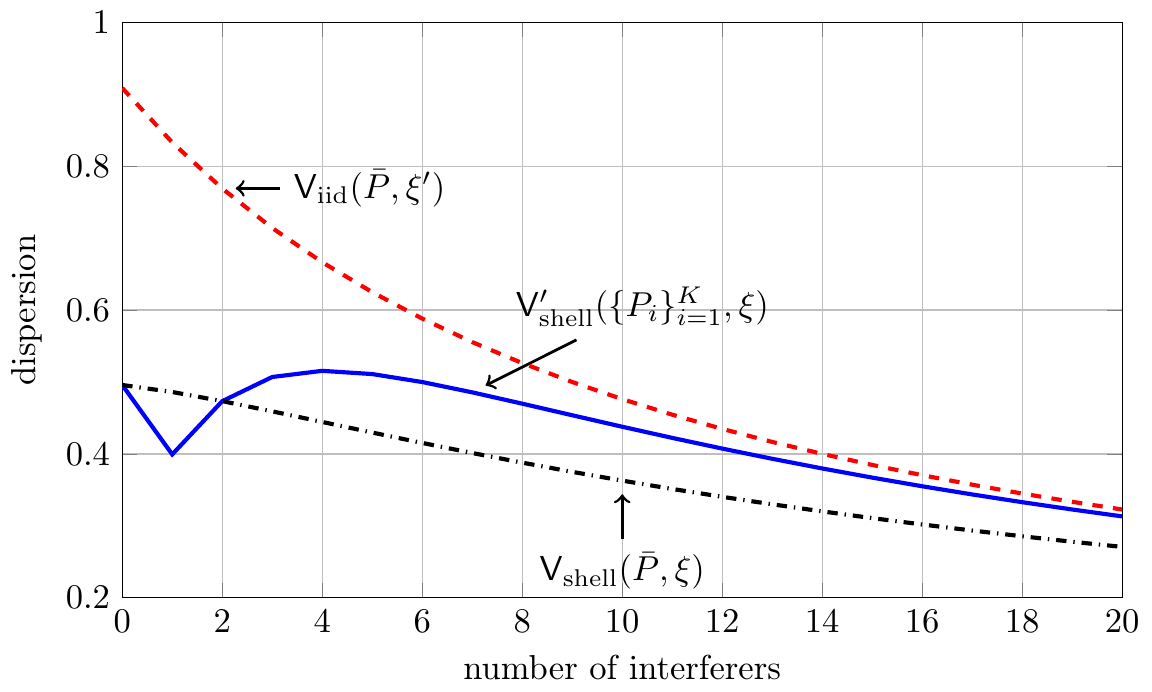}
  \caption{Dispersion as a function of the number of interferers $K-1$ for both shell and \iid codes. 
  Here, $P_1=10$, $P_2=\dots=P_K=1$ and $\bar{P}=P_1/(1+\sum_{i=2}^K P_i)$. The noise $Z^n$ is \iid Gaussian with unit variance.} 
  \label{fig:figs_dispersion_int} 
\end{figure}

\subsection{Numerical Illustration}
In~Fig.~\ref{fig:figs_dispersion_int}, we plot both $\rvV_{\mathrm{shell}}'( \{P_j\}_{j=1}^K ,\xi)$ and $\rvV_{\mathrm{iid}} (\bar P,\xi')$ for the case $P_1=10$, $P_2=\ldots=P_K=1$. 
The noise $Z^n$ is \iid Gaussian with unit variance ($\xi=3$).
\purple{We recall that in the shell case, although the noise is \iid Gaussian, the noise-plus-interference term is neither Gaussian nor \iid}
For comparison, we also plot $\rvV_{\mathrm{shell}}( \bar{P},\xi)$, which corresponds to the dispersion for the case when the intended sender uses a shell code, whereas the interferers use  \iid codes. 
The resulting dispersion, which follows directly from Theorem~\ref{thm:disp}, takes a particularly simple form, namely, the Gaussian dispersion~\eqref{eq:gaussian_dispersion} computed at the SINR $\bar{P}$. 
As illustrated in Fig.~\ref{fig:figs_dispersion_int}, there is no ordering between $\rvV_{\mathrm{shell}}( \bar{P},\xi)$ and $\rvV_{\mathrm{shell}}'( \{P_j\}_{j=1}^K ,\xi)$. 
\purple{As shown if Fig.~\ref{fig:figs_dispersion_int}, for the case $P_1=10$, $P_2=\ldots=P_K=1$} shell interference is preferable \blue{(i.e., gives a smaller dispersion to user 1)} when there is a single interferer, but \iid~interference is preferable when there are three or more users.


\section{Proof of  Theorem \ref{thm:disp}}   \label{sec:prf_add}
Our analysis makes  extensive use of the ``mismatched'' information density
\begin{align} 
\tilde{\iota}(x,y) &:= \log\frac{ \calN( y;x,1)}{ \calN( y; 0,P+1) } \label{eqn:info_dens0}\\
&= \rvC(P) + \frac{y^2}{2(P+1)}-\frac{(y-x)^2}{2}. \label{eqn:info_dens}
\end{align}
This is the information density of the Gaussian channel $\calN( y;x,1)$; indeed, the denominator in~\eqref{eqn:info_dens0} is its capacity-achieving output distribution $\calN( y; 0,P+1)$.  We define 
\begin{equation}
\tilde{\iota}^n(\bx,\by) :=\sum_{i=1}^n \tilde{\iota}(x_i,y_i).
\end{equation}
\subsection{Proof of \eqref{eqn:asymp_shell} in Theorem \ref{thm:disp}}

\subsubsection{Proof of the Direct Part of \eqref{eqn:asymp_shell}} \label{sec:direct}
We shall obtain a lower bound on $\log M^*_{\mathrm{shell}}(n,\eps,P;\rmP_Z)$, whose asymptotic expansion coincides with the right-hand side of~\eqref{eqn:asymp_shell}. 
 By \eqref{eqn:info_dens}, the NN rule is equivalent to maximizing $\tilde{\iota}^n(\bx(w),\by)$ over $w\in [1:M]$.  Hence, by using a ``mismatched'' version of the   random coding union bound~\cite[Th.~16]{PPV10}, we conclude that the ensemble error probability $\barp_{\rme, n} $ can be bounded as 
\begin{align}\label{eq:RCU}
& \barp_{\rme, n}  \le \bbE \Big[ \min\bigl\{ 1, \nn\\*
 &\qquad M\Pr ( \tilde{\iota}^n(\barX^n,Y^n)\ge \tilde{\iota}^n(X^n,Y^n) |X^n,Y^n)  \bigr\}\Big],
\end{align}
where $(\barX^n,X^n,Y^n)\sim \rmf_{X^n}^{({\mathrm{shell}})}(\bar{\bx})   \rmf_{X^n}^{({\mathrm{shell}})}(\bx) \rmP_{Y^n|X^n}(\by|\bx)$.
Defining 
\begin{IEEEeqnarray}{rCL}\label{eq:def_g}
  g(t,\by):=\Pr\lefto[\tilde{\iota}^n(\barX^n,\by)\geq t\right],
\end{IEEEeqnarray}
we can rewrite~\eqref{eq:RCU} as
\begin{IEEEeqnarray}{rCL}\label{eq:RCU-g}
  \barp_{\rme, n} \le \bbE \lefto[ \min\bigl\{ 1,M g(\tilde{\iota}^n(X^n,Y^n),Y^n) \bigr\}\right].
\end{IEEEeqnarray}
For sufficiently large $n$, the function $g(t,\by)$ can be bounded as follows:
\begin{IEEEeqnarray}{rCL}
  g(t,\by)&\leq& \frac{K_0 e^{-t}}{\sqrt{n}} \label{eq:vincent} \\
  &\leq& K_0 e^{-t}. \label{eq:vincent_2}
\end{IEEEeqnarray}
Here, $K_0$ is a finite constant that does not depend on $\by$. 
The inequality~\eqref{eq:vincent} is given in~\cite[Eq.~(58)]{TanTom13a} (recall that $\tilde{\iota}^n(\barX^n,\by)$ is the information density of an additive \emph{Gaussian} channel), and~\eqref{eq:vincent_2} follows because $\sqrt{n}\geq 1$.
\textcolor{black}{We note that one can also upper-bound $g(t,\by)$ by the expression in~\eqref{eq:vincent_2} directly using~\cite[Prop.~2]{Mol13c} (see also~\cite[p.~2317]{PPV10}).}

Substituting~\eqref{eq:vincent_2} into~\eqref{eq:RCU-g} and using that for every real-valued random variable \textcolor{black}{$J$} and every positive integer $n$
\begin{IEEEeqnarray}{rCL}
  \bbE\bigl[\min\lefto\{1,J\right\}\bigr]&\leq& \Pr\lefto[J>\frac{1}{\sqrt{n}}\right] +\frac{1}{\sqrt{n}}\Pr\lefto[J\leq \frac{1}{\sqrt{n}}\right] \\
  &\leq&  \Pr\lefto[J>\frac{1}{\sqrt{n}}\right] +\frac{1}{\sqrt{n}},
\end{IEEEeqnarray}
we further upper-bound the right-hand-side of~\eqref{eq:RCU-g} as follows:
\begin{align}
&  \barp_{\rme, n} \nn\\*
  &\le  \Pr\lefto[MK_0 e^{-\tilde{\iota}^{n}(X^n,Y^n)} \geq \frac{1}{\sqrt{n}}\right]+\frac{1}{\sqrt{n}}\\
  &\leq  \Pr\biggl[ \log M - \Big( n\rvC(P) + \frac{\|Y^n\|^2}{2(P+1)}- \frac{\|Y^n-X^n\|^2}{2}\Big) \nn\\*
  & \quad \hspace*{3.5cm} \ge -\log\bigl(K_0\sqrt{n}\bigr) \bigg]+\frac{1}{\sqrt{n}}\label{eqn:upper_bd_pe0}\\
  & =\Pr\biggl[\frac{\|Z^n\|^{2}}{2} - \frac{\|X^n + Z^n\|^{2}}{2(P+1)}\nn\\*
  &\quad \hspace*{1.2cm} \ge n\rvC(P) -  \log M-\log\bigl(K_0\sqrt{n}\bigr)\bigg]+\frac{1}{\sqrt{n}}  \label{eqn:upper_bd_pe}\\
  &=\Pr\biggl[(P+1) \|Z^n\|^2 - \|X^n+Z^n\|^2 \nn\\*
  &\quad\ge 2(P+1) \Big( n \rvC(P)-\log M-\log\bigl(K_0\sqrt{n}\bigr)\Big) \biggr]+\frac{1}{\sqrt{n}} \label{eqn:upper_bd_pe_2}\\
   & =\Pr\biggl[P\|Z^n\|^{2}-nP-2\langle X^n,Z^n\rangle \nn\\*
   &\quad
   \ge 2(P+1)\Big(n\rvC(P)-\log M-\log\bigl(K_0\sqrt{n}\bigr)\Big)\bigg]+\frac{1}{\sqrt{n}}.\label{eq:RCUs2}
\end{align}
%
%
%
%
Here, \eqref{eqn:upper_bd_pe0} follows from~\eqref{eqn:info_dens}, \eqref{eqn:upper_bd_pe_2} follows since $Y^n = X^n + Z^n$, and \eqref{eq:RCUs2}
follows because $\|X^n\|^{2}=nP$ almost surely. 
Now we make use of the Berry-Esseen Theorem for functions of random vectors in Theorem \ref{thm:berry}  stated in Appendix \ref{app:be_func} (which is based on \textcolor{black}{\cite[Prop.~1]{Mol13c} and~\cite[Prop.~1]{Iri15}}). By proceeding similarly as in~\cite[Sec.~IV.D]{Mol13c}. 
A shell codeword  can be written as 
\begin{equation}
X^n = \sqrt{nP} \frac{\tilX^n}{\|\tilX^n\|},
\end{equation}
where $\tilX^n\sim\calN(\mathbf{0},\bI_n)$. We can write  $P\|Z^n\|^{2}-nP-2\langle X^n,Z^n\rangle$  in terms of the zero-mean \iid random variables 
\begin{align}
A_{1,i} &:= 1-Z_i^2 , \label{eqn:defAi1} \\ A_{2,i} & := \sqrt{P}\tilX_i Z_i,  \label{eqn:defAi2}\\ A_{3,i}& :=  \tilX_i^2 - 1, \label{eqn:defAi3}
\end{align}
for $i=1,\dots,n$, and the smooth function
\begin{equation}
f(a_1, a_2,a_3) := Pa_1 + \frac{2a_2}{\sqrt{1+a_3}}.
\end{equation}
Specifically, we have
\begin{multline}
-n f\lefto(\frac{1}{n}\sum_{i=1}^n A_{1,i}, \frac{1}{n}\sum_{i=1}^n A_{2,i}, \frac{1}{n}\sum_{i=1}^n A_{3,i}  \right) \\
= P\|Z^n\|^{2}-nP-2\langle X^n,Z^n\rangle. \label{eqn:relatef}
\end{multline}
 Now, the Jacobian matrix of $f$ evaluated at $\mathbf{0}$ is $\bJ = [P\,\,\, 2\,\,\, 0]$ and the covariance matrix $\bV$ of the vector $[A_{1,1}\,\,\, A_{2,1} \,\,\, A_{3,1} ]$ is $\bV=\mathrm{diag}( [ \xi-1 \,\,\, P\,\,\, 2 ]^T )$.
\blue{We now apply the Berry Esseen theorem for functions in Theorem \ref{thm:berry} in Appendix \ref{app:be_func}.   
  With the above notation, Theorem \ref{thm:berry} states that $(P\|Z^n\|^{2}-nP-2\langle X^n,Z^n\rangle)/\sqrt{n}$ converges in distribution to a univariate, zero-mean normal random variable with variance $\bJ \bV \bJ^T = P^2(\xi-1 ) + 4P$, and that the convergence rate is $O(1/\sqrt{n})$.} We thus conclude that
\begin{multline}
\barp_{\rme, n}  \leq \rmQ\lefto( \frac{2(P+1)  \big( n\rvC(P) -\log M -\log\bigl(K_0\sqrt{n}\bigr) \big) }{\sqrt{n (P^2 (\xi-1) + 4P ) } } \right)  \\
\qquad\qquad+O\Big(\frac{1}{\sqrt{n}} \Big). \label{eqn:q_fun}
\end{multline}
Equating the right-hand-side of~\eqref{eqn:q_fun} to $\eps$, solving for $\log M$ and performing  a first-order Taylor expansion of $\rmQ^{-1}(\cdot)$ about $\eps$, we establish the lower bound corresponding to \eqref{eqn:asymp_shell}--\eqref{eqn:disp_shell}.

\subsubsection{Proof of the Ensemble Tightness Part of \eqref{eqn:asymp_shell}} \label{sec:prf-she}
Since the probability of ties  for the NN rule in~\eqref{eqn:NN} is zero,  the exact random coding probability can be written as 
\begin{equation}
\barp_{\rme, n} = \bbE [   \barp_{\rme, n} (X^n,Y^n) ], \label{eqn:exact_prob1}
\end{equation}
where   
\begin{equation}
\barp_{\rme, n} (\bx,\by) = 1-\left( 1- \Pr [ \|\by - \barX^n \|\le \|\by-\bx\|  ] \right)^{M-1} \label{eqn:exact_prob2}
\end{equation}
and $(X^n,\barX^n,Y^n)$ are distributed   as in  Section~\ref{sec:direct}.
Let $\bz:=\by-\bx$.
The probability in \eqref{eqn:exact_prob2} depends on $(\bx,\by)$  only through the powers $\hatP_Y =  \|\by\|^2/n$ and $\hatP_Z = \|\bz\|^2/n$.
Indeed, since $\|X^n\|^2 = nP$ almost surely, we have
\begin{align}
&  \Pr \lefto[ \| \barX^n - \by\|^2 \le\|\bz\|^2\right] \nn\\*
  &=\Pr \lefto[ \|\barX^n\|^2 - 2  \langle \barX^n,\by\rangle +  \|\by\|^2\le\|\bz\|^2 \right]\label{eqn:start_pt0} \\
  & =\Pr\lefto[ 2\langle \barX^n,\by\rangle\ge n \hatP_Y + nP - n\hatP_Z\right]\label{eqn:start_pt}\\
  &=\Pr\lefto[ \barX_1 \ge\frac{n \hatP_Y + nP - n\hatP_Z}{ 2\sqrt{n \hatP_Y}}\right].
\end{align}
Here, $\barX_1$ denotes the first symbol in the vector $\barX^n$; the last step follows because $\barX^n$ is spherically symmetric, and, hence, we can transform $\by$  into $(\sqrt{n\hat{P}_Y},0\dots,0)$ by performing a rotation that does not change the probability.
For convenience, we set
\begin{IEEEeqnarray}{rCL}
  \Psi(\hatP_Y,\hatP_Z):= \Pr\lefto[ \barX_1 \ge\frac{n \hatP_Y + nP - n\hatP_Z}{ 2\sqrt{n \hatP_Y}}\right] \label{eqn:Psi_def}
\end{IEEEeqnarray}
so that
\begin{IEEEeqnarray}{rCL}\label{eq:pe_converse_alt}
  \barp_{\rme, n} (\bx,\by) = 1-\left( 1- \Psi(\hatP_Y,\hatP_Z)
  \right)^{M-1}.
\end{IEEEeqnarray}
\textcolor{black}{Because of \eqref{eqn:lap}, we shall assume without loss of generality that   $\liminf_{n\to\infty}(\log M)/n>0$, i.e., that $M$ grows (at least) exponentially with the blocklength. }
Let 
\begin{align}
c_z&:=\xi-1, \quad\mbox{and}\\ c_y&:=\xi-1+4P .  \label{eqn:defcy}
\end{align}
We define the {\em typical sets of powers}
\begin{align}
\calP_Y  := \left\{ \hatp_Y \in\bbR : |\hatp_Y - (P+1) |\le \sqrt{c_z\frac{\log n}{n}}\right\}\label{eqn:typ1} 
\end{align}
and
\begin{align}
\calP_Z  := \left\{ \hatp_Z \in\bbR : |\hatp_Z - 1 |\le  \sqrt{c_y\frac{\log n}{n}}\right\} .\label{eqn:typ2}
\end{align}
Moreover, we fix $\eta \in (0,2P)$, and define the following additional typical set to ensure that $\hatp_Y+P-\hatp_Z$ is uniformly bounded away from zero: 
\begin{equation}
\calQ := \left\{(\hatp_Y ,\hatp_Z)\in\bbR^2: \hatp_Y+P-\hatp_Z >\eta \right\}.\label{eqn:typ3}
\end{equation}
For brevity, we combine the above typical sets into one:
\begin{equation}
\calT:= (\calP_Y\times\calP_Z)\cap\calQ.  \label{eqn:defT}
\end{equation}
In Appendix \ref{app:moderate}, we show using  the Berry Esseen theorem for functions of random vectors (Theorem \ref{thm:berry} in Appendix \ref{app:be_func}) and the assumption that $\bbE [Z^6]$ is finite that
\begin{align}
\Pr\biggl[\Big(\frac{1}{n}\|Y^n\|^2,\frac{1}{n}\|Z^n\|^2\Big)\notin\calT \bigg] = O\Big(\frac{1}{\sqrt{n}} \Big) . \label{eqn:modZ}
\end{align}
The implied constant  in the $O(\fndot)$ notation  above depends  only on $\xi, \bbE[Z^6]$ and $P$. 

Now consider an arbitrary pair of powers $(\hatP_Y,\hatP_Z)$ such that $\Psi(\hatP_Y,\hatP_Z)\ge{n}/{(M-1)}$. In this case, we have
\begin{align}
& 1- \left(1-\Psi(\hatP_Y,\hatP_Z)  \right)^{M-1}\nn\\*
 & \ge 1-\left( 1-\frac{n}{M-1}\right)^{M-1}\\
&= 1-\left( \Big( 1-\frac{n}{M-1}\Big)^{\frac{M-1}{n}} \right)^n\\
& = 1-\left( e^{-1} (1+o(1))\right)^n \\
&\ge 1-e^{-n\gamma} \label{eqn:bound_Psi}
\end{align}
where we used the fact that $n/(M-1)\to 0$ (recall that we assumed that \textcolor{black}{$\liminf_{n\to\infty}(\log M)/n>0$}) and also that $(1-\zeta^{-1})^\zeta\to {e^{-1}}$ as $\zeta\to \infty$.
The inequality~\eqref{eqn:bound_Psi} holds for any given $\gamma\in (0,1)$ provided that $n$ is sufficiently large.

Combining the analyses in the previous two paragraphs, we obtain from \eqref{eqn:exact_prob1} and \eqref{eq:pe_converse_alt} that 
\begin{align}
&\barp_{\rme, n}\nn\\*
& \ge\Pr\biggl[  \Psi(\hatP_Y,\hatP_Z)  \ge \frac{n}{M-1}  \cap (\hatP_Y,\hatP_Z) \in  \calT\bigg] (1-e^{-n\gamma }) \\
&= \Pr\biggl[  \Psi(\hatP_Y,\hatP_Z)  \ge \frac{n}{M-1}  \cap (\hatP_Y,\hatP_Z) \in  \calT\bigg] + O\Big(\frac{1}{\sqrt{n}} \Big). \label{eqn:lower_bd}
\end{align}
Ensemble tightness will be established if we can show  that for all typical powers $(\hatP_Y,\hatP_Z) \in \calT$,
\begin{align}
   \Psi(\hatP_Y,\hatP_Z)  \ge  q(n)\exp\biggl(   -n \Big( \rvC(P)   +  \frac{\hatP_Y}{2(P  +  1)}   -  \frac{\hatP_Z}{2}\Big)\bigg) \label{eqn:tightness_ens}
\end{align} 
for some $q(n)$ satisfying $\log q(n)=O(\log n)$.  
Indeed, if~\eqref{eqn:tightness_ens} holds, we can further lower-bound~\eqref{eqn:lower_bd} as 
\begin{align}
& \barp_{\rme, n} + O\Big(\frac{1}{\sqrt{n}} \Big)  \nn\\*
 &\ge\Pr\biggl[ -n\Big( \rvC(P) +\frac{\hatP_Y}{2(P+1)} -\frac{\hatP_Z}{2}\Big)\nn\\*
& \quad \ge -\log M +  O(\log n) \cap (\hatP_Y,\hatP_Z) \in  \calT\bigg]\label{eq:ub_first_step}\\
&=\Pr\biggl[\rvC(P)   +  \frac{\hatP_Y}{2(P+1)}   -  \frac{\hatP_Z}{2}  \le     \frac{\log M}{n}  +   O\Big(\frac{\log n}{n}\Big)\bigg]   \label{eqn:usetyp}\\
&=\Pr\biggl[ \frac{\|Z^n\|^2}{2}-\frac{\|Y^n\|^2}{2(P+1)}\ge n \rvC(P) - \log M +O(\log n) \bigg] \label{eqn:apply_defs_typ}\\
&=\Pr\biggl[P\|Z^n\|^{2}-nP-2\langle X^n,Z^n\rangle \nn\\*
&\quad  \ge 2(P+1)\Big(n\rvC(P)-\log M+O(\log n)\Big)\bigg]
\end{align}
In~\eqref{eq:ub_first_step} we used that $\log (M-1)\geq \log M - \log 2$ (assuming without loss of generality $M>1$),
in~\eqref{eqn:usetyp} we used that $\Pr[\calA \cap \calB]\geq \Pr[\calA]+\Pr[\calB]-1$ and the concentration bounds of the empirical powers in the typical set in~\eqref{eqn:modZ}, and in \eqref{eqn:apply_defs_typ} we applied the definitions of $\hatP_Y$ and $\hatP_Z$. 
To complete the proof (\emph{sans} the justification of~\eqref{eqn:tightness_ens}),
we follow steps~\eqref{eq:RCUs2}--\eqref{eqn:q_fun} in the direct part, which are all tight in  the second-order sense. 

  Now we prove \eqref{eqn:tightness_ens}. 
We fix $\delta>0$ 
and bound \eqref{eqn:Psi_def} as
\begin{align}
&\Psi(\hatP_Y,\hatP_Z)\nn\\*
 &= \Pr\lefto[ \barX_1 \ge\frac{n \hatP_Y + nP - n\hatP_Z}{ 2\sqrt{n \hatP_Y}}\right]\\
 & \ge\Pr\biggl[\frac{n\hat{P}_{Y} +  nP -  n\hat{P}_{Z}}{2\sqrt{n\hat{P}_{Y}}} \le \barX_1 \nn\\*
&\qquad \le \frac{n\hat{P}_{Y} + nP -  n\hat{P}_{Z}}{2\sqrt{n\hat{P}_{Y}}}\Big(1 + \frac{\delta}{n}\Big)\bigg]\label{eq:Completion5}\\
 & \ge\frac{\delta(\hat{P}_{Y}+P-\hat{P}_{Z})}{2\sqrt{n\hat{P}_{Y}}}\min_{x}f_{\barX_1}(x),\label{eq:Completion6}
\end{align}
where the minimization in \eqref{eq:Completion6} is over the interval in \eqref{eq:Completion5}, and  $f_{\barX_1}$ is the probability density function of $\barX_1$, which is given by~\cite[Eq.\ (4)]{Stam}:
\begin{equation}
f_{\barX_1}(x)=\frac{1}{\sqrt{\pi nP}}\frac{\Gamma(\frac{n}{2})}{\Gamma(\frac{n-1}{2})}\bigg(1-\frac{x^{2}}{nP}\bigg)^{{(n-3)}/{2}}\mathbf{1}\big\{ x^{2}\le nP\big\}.\label{eq:DPC_ShellDistr}
\end{equation}
Note that since the right-hand side of~\eqref{eq:DPC_ShellDistr} is a decreasing function of $|x|$, we can further lower-bound~\eqref{eq:Completion6} as follows:
\begin{align}
&  \Psi(\hatP_Y,\hatP_Z)  \nn\\*
&\ge  \frac{\delta(\hat{P}_{Y}+P-\hat{P}_{Z})}{2\sqrt{n\hat{P}_{Y}}}\frac{1}{\sqrt{\pi nP}}\frac{\Gamma(\frac{n}{2})}{\Gamma(\frac{n-1}{2})} \nn\\*
  &\,\,\times\left[1-\frac{1}{nP}\Bigg(\frac{n\hat{P}_{Y}+nP-n\hat{P}_{Z}}{2\sqrt{n\hat{P}_{Y}}}\bigg(1+\frac{\delta}{n}\bigg)\Bigg)^{2}\right]^{(n-3)/2}. \label{eqn:psi_lb}
\end{align}
Since $\hat{P}_{Y}$
and $\hat{P}_{Z}$ are bounded within the typical sets in \eqref{eqn:typ1}--\eqref{eqn:typ2},  we have that for all sufficiently large $n$ and for some constant $K_1$ (depending only on $P$  and $\delta$) that
\begin{equation}\label{eq:the_actual_term_to_bound}
\Bigg(\frac{n\hat{P}_{Y}+nP-n\hat{P}_{Z}}{2\sqrt{n\hat{P}_{Y}}}\bigg(1+\frac{\delta}{n}\bigg)\Bigg)^{2}\le\frac{n(\hat{P}_{Y}+P-\hat{P}_{Z})^{2}}{4\hat{P}_{Y}}+K_1.
\end{equation}
Moreover, ${\Gamma(\frac{n}{2})}/{\Gamma(\frac{n-1}{2})}$
behaves as $\Theta(\sqrt{n})$, and thus the first three factors in \eqref{eqn:psi_lb} (the first of which is positive because $(\hatP_Y,\hatP_Z)\in\calT$ implies that $\hatP_Y+P-\hatP_Z>\eta>0$) can be combined into a single prefactor
$p_{0}^{\prime}(n)$ satisfying $\log p_{0}^{\prime}(n)=O(\log n)$.
Hence,
\begin{align}
&\Psi(\hat{P}_{Y},\hat{P}_{Z}) \nn\\*
& \ge p_{0}^{\prime}(n)\bigg(1-\frac{(\hat{P}_{Y}+P-\hat{P}_{Z})^{2}}{4P\hat{P}_{Y}}-\frac{K_1}{nP}\bigg)^{{(n-3)}/{2}}\\
 & \geq p_{0}^{\prime}(n)\bigg(\bigg(1-\frac{(\hat{P}_{Y}+P-\hat{P}_{Z})^{2}}{4P\hat{P}_{Y}}\bigg)\Big(1-\frac{K_2}{nP}\Big)\bigg)^{{(n-3)}/{2}}\label{eq:Completion10}\\
 & \ge p_{0}^{\prime\prime}(n)\bigg(1-\frac{(\hat{P}_{Y}+P-\hat{P}_{Z})^{2}}{4P\hat{P}_{Y}}\bigg)^{{n}/{2}}\label{eq:Completion11}\\
 & =p_{0}^{\prime\prime}(n)\exp\Biggl(\frac{n}{2}\log\bigg(1-\frac{(\hat{P}_{Y}+P-\hat{P}_{Z})^{2}}{4P\hat{P}_{Y}}\bigg)\Bigg).\label{eq:Completion12}
\end{align}
Here,~\eqref{eq:Completion10} holds for some finite constant $K_2$. 
and \eqref{eq:Completion11} follows by absorbing additional terms in a new prefactor $p_{0}^{\prime\prime}(n)$ still satisfying $\log p_{0}^{\prime\prime}(n) =O(\log n)$; specifically, this step uses again \eqref{eqn:typ1}--\eqref{eqn:typ2}, as well as the limit
\begin{equation}
\lim_{n\to \infty}\Big(1-\frac{K_2}{nP }\Big)^{n/2}= e^{{-K_2}/{(2P)}},
\end{equation}
which is a constant.

We prove \eqref{eqn:tightness_ens} by performing the following Taylor expansion of $\frac{1}{2}\log\bigl(1- {(\hat{P}_{Y}+P-\hat{P}_{Z})^{2}}/{(4P\hat{P}_{Y})}\big)$ in~\eqref{eq:Completion12}
about $(\hat{P}_{Y},\hat{P}_{Z})=(P+1,1)$:
\begin{align}
 & -\frac{1}{2}\log\Bigl(1-\frac{(\hat{P}_{Y}+P-\hat{P}_{Z})^{2}}{4P\hat{P}_{Y}}\Big)\nonumber \\
 &  =\frac{1}{2}\log(P+1)+\frac{1}{2(P+1)}\Big(\hat{P}_{Y}-(P+1)\Big)-\frac{1}{2}\Big(\hat{P}_{Z}-1\Big)\nn\\*
&\qquad  +O\Bigl(\big|\hat{P}_{Y}-(P+1)\big|^{2}+\big|\hat{P}_{Z}-1\big|^{2}\Big)\\
 &  =\rvC(P)+\frac{\hat{P}_{Y}}{2(P+1)}-\frac{\hat{P}_{Z}}{2}+O\Big(\frac{\log n}{n}\Big).
\end{align}
Here, the remainder term is $O\big( {(\log n)}/{n}\big)$ due to the
definition of the typical sets in \eqref{eqn:typ1}--\eqref{eqn:typ2}.
This remainder term can be factored into the  prefactor
in \eqref{eq:Completion12}, yielding $q(n)$ in~\eqref{eqn:tightness_ens}. 
The proof of \eqref{eqn:tightness_ens} is
thus complete.

\subsection{Proof of \eqref{eqn:asymp_iid} in Theorem \ref{thm:disp}}
\subsubsection{Proof of the Direct Part of  \eqref{eqn:asymp_iid}}
Since $X^n$ is \iid, the direct part of~\eqref{eqn:asymp_iid} can be established using the standard Berry-Esseen central limit theorem, and following the steps detailed in~\cite[Sec. IV.A]{PPV10}.
We just need to verify that  the expectation and the variance of $\tilde{\iota}(X,Y)$ (when $X\sim \calN(0,P)$ and $Y=X+Z$)  are $\rvC(P)$ and $\rvV_{\mathrm{iid}}(P,\xi)$, respectively. 
The former is straightforward, and for the latter,  using the statistics of $Z$ in \eqref{eqn:moments} and the fact that $X$ is independent of $Z$, one has 
\begin{align}
&\var\lefto[ \tilde{\iota}(X,Y)  \right] \nn\\*
& =\var\lefto[ \frac{(X+Z)^2}{2(P+1) } - \frac{Z^2}{2}\right] \\
&=\frac{1}{4(P+1)^2} \var[ X^2 +2XZ - PZ^2 ]\\
&=\frac{1}{4(P+1)^2} \Big(\bbE [ (X^2 +2XZ - PZ^2)^2 ] \nn\\*
&\qquad  -\bbE [  X^2 +2XZ - PZ^2  ]^2  \Big) \\
&=\frac{1}{4(P+1)^2} \bbE [  X^4  +4X^2 Z^2 + P^2 Z^4 - 2X^3Z \nn\\*
&\qquad-2PX^2 Z^2-2PXZ^3+4X^3Z-4PXZ^3]\\
&=\rvV_{\mathrm{iid}}(P,\xi),
\end{align}
where in the last step we applied $\bbE[X^3Z] = \bbE[XZ^3] = 0$ along with standard Gaussian moments.
This completes the  proof of the direct part of~\eqref{eqn:asymp_iid}.

\subsubsection{Proof of the Ensemble Tightness Part of  \eqref{eqn:asymp_iid}}
We prove only \eqref{eqn:tightness_ens}, since all other steps (including the proof of~\eqref{eqn:modZ}) follow those in Section~\ref{sec:prf-she} along with standard steps in dispersion analyses \cite{PPV10}. 
We again
start with the left-hand-side of \eqref{eqn:start_pt0}, \blue{and recall that the probability remains unchanged when we transform $\by$ arbitrarily subject to its power remaining unchanged.  This time, we choose the form
$\big(\sqrt{\hat{P}_{Y}},\cdots,\sqrt{\hat{P}_{Y}}\big)$, thus yielding}
\begin{align}
\Psi(\hat{P}_{Y},\hat{P}_{Z}) & =\Pr\lefto[\sum_{i=1}^{n}\Big(\overline{X}_{i}-\sqrt{\hat{P}_{Y}}\Big)^{2}\le n\hat{P}_{Z}\right]\\
 & =\Pr\lefto[-\frac{1}{P}\sum_{i=1}^{n}\Big(\overline{X}_{i}-\sqrt{\hat{P}_{Y}}\Big)^{2}\ge-n\frac{\hat{P}_{Z}}{P}\right]\label{eq:TightIID2}
\end{align}
Here, we consider $(\hat{P}_{Y},\hat{P}_{Z})$ belonging to the typical set $\calT$ defined in \eqref{eqn:defT}. Since $\overline{X}_{i}\sim \calN(0,P)$, each term $\big(\overline{X}_{i}-\sqrt{\hat{P}_{Y}}\big)^{2}/P$
follows a non-central $\chi^{2}$-distribution with one degree of freedom
and noncentrality parameter ${\hatP_{Y}}/{P}$. 
Its moment generating function is thus \cite[Sec.\ 2.2.12]{Tanizaki}
\begin{align}
M(s) &:= \bbE \lefto[ \exp\lefto(s\cdot  \frac{1}{P}\Big(\overline{X}_{i}-\sqrt{\hat{P}_{Y}}\Big)^{2}  \right)  \right] \\ &=\frac{1}{\sqrt{1-2s}}\exp\lefto(\frac{\hatP_{Y}s}{P(1-2s)}\right).
\end{align}
Now using the   strong large deviations (exact asymptotics) theorem by Bahadur and Ranga Rao~\cite{BahRR60} (see also  \cite[Theorem 3.7.4]{Dembo} and \cite{CS93}), we observe that 
\begin{equation}
\Psi(\hat{P}_{Y},\hat{P}_{Z})=q(n)\exp\bigl(-nE (\hat{P}_{Y},\hat{P}_{Z}) \big)
\end{equation}
for some function $q(n)$ satisfying $ q(n) =\Theta({1}/{\sqrt{n}})$, where the exponent is
\begin{align}
E(\hat{P}_{Y},\hat{P}_{Z}) &:=\sup_{s\ge 0} \left\{-\log M(-s) -s\frac{\hatP_Z}{P}\right\}\\
&=\sup_{s\ge 0} \left\{ \frac{\hatP_Ys}{P(1+2s)}+\frac{1}{2}\log(1+2s)-s\frac{\hatP_Z}{P}\right\}. \label{eqn:sup_s}
\end{align}
The objective function in \eqref{eqn:sup_s} is strictly concave. By a direct differentiation, we find that the supremum is achieved by
\begin{equation}
s^{*}=\frac{P-2\hatP_{Z}+\sqrt{P^{2}+4\hatP_{Y}\hatP_{Z}}}{4\hatP_{Z}}.
\end{equation}
Hence,
\begin{align}
&E(\hat{P}_{Y},\hat{P}_{Z}) =\frac{1}{2}\log\bigg(1+\frac{P-2  \hatP_{Z}+\sqrt{P^{2} 
 +4 \hatP_{Y} \hatP_{Z}}}{2\hatP_{Z}}\bigg) \nn\\*
 &\qquad+\frac{ \hatP_{Y}(P-2\hatP_{Z}+\sqrt{P^{2}+4\hatP_{Y}\hatP_{Z}}}{4P\hatP_{Z}\Big(1+\frac{P-2\hatP_{Z}+\sqrt{P^{2}+4\hatP_{Y}\hatP_{Z}}}{2\hatP_{Z}}\Big)}\nn\\*
 &\qquad-\frac{P-2\hatP_{Z}+\sqrt{P^{2}+4\hatP_{Y}\hatP_{Z}}}{4P}.
\end{align}
Performing a Taylor expansion of $E(\hat{P}_{Y},\hat{P}_{Z})$ about $(\hat{P}_{Y},\hat{P}_{Z})=( P+1,1)$ similarly to the shell case in Section~\ref{sec:prf-she}, and using  \eqref{eqn:typ1}--\eqref{eqn:typ2}, we obtain
\begin{equation}
E(\hat{P}_{Y},\hat{P}_{Z}) =\rvC(P)+\frac{\hat{P}_{Y}}{2(P+1)}-\frac{\hat{P}_{Z}}{2}+O\Big(\frac{\log n}{n}\Big),
\end{equation}
thus completing the proof of \eqref{eqn:tightness_ens}.

\section{Proof of  Theorem \ref{thm:disp2}}   \label{sec:prf_add2}
We shall first present the proof of the \iid interference case~\eqref{eqn:interf_iid} and then move to the shell interference case~\eqref{eqn:interf_shell}.

\subsection{Proof of \eqref{eqn:interf_iid} in Theorem \ref{thm:disp2}} \label{sec:prf_inter_iid}
As all the users employ \iid Gaussian codebooks, the asymptotic expansion~\eqref{eqn:interf_iid} follows directly from~\eqref{eqn:asymp_iid} in Theorem~\ref{thm:disp}, provided that the total noise variance is normalized to one.  
Specifically, since the codewords of users $2,\ldots, K$ are to be treated as noise, user $1$ observes effective noise 
\begin{equation}
\tilZ:=X_2 + \ldots + X_K+Z.
\end{equation}
Recall that $\tilP :=\sum_{i=2}^K P_i$ is the sum power of the interfering codewords. 
The mutual independence of $X_2,\ldots, X_K, Z$ and the fact that $\bbE [ X_i]=0, i \in [ 2 : K]$ imply that
\begin{equation}
\bbE[ \tilZ^2 ] =\tilP+1. 
\end{equation}
Let 
$
Z_0:= X_2 + \ldots + X_K.
$ 
  The \textcolor{black}{fourth moment of $\tilZ$} can be computed as follows:
%
\begin{IEEEeqnarray}{rCL}
  \bbE[\tilZ^4]&=&\bbE[(Z_0+Z)^4]\\
  &=& 3\tilP^2 +6\tilP+\xi.
\end{IEEEeqnarray}

Now in order to apply \eqref{eqn:asymp_iid} in Theorem \ref{thm:disp}, we need to normalize the channel input-output relation, so as to make the effective noise of unit variance. 
We do this by setting $\barZ := \tilZ / \sqrt{\tilP+1}$, and by rescaling the power of the intended user to $\barP={P_{1}}/{(\tilP+1)}$  (see~\eqref{eqn:sinr}).
Note that
\begin{equation}
\bbE\bigl[\barZ^4\bigr]=\frac{3\tilP^2 +6\tilP+\xi}{\big( \tilP+1\big)^{2}}.
\end{equation}
Substituting into \eqref{eqn:disp_iid} with the identifications $P\equiv\barP$ and $\xi\equiv\bbE\bigl[\barZ^4\bigr]$, we obtain~\eqref{eqn:interf_iid}--\eqref{eqn:xi_prime}.


\subsection{Proof of \eqref{eqn:interf_shell} in Theorem \ref{thm:disp2}}
The proof follows along the same lines as the proof of \eqref{eqn:disp_shell} in Section~\ref{sec:prf_add}.
Hence, we only provide the details of the changes that need to be made. 

 Let $\widetilde{X}^n=X_2^n +\ldots+X_K^n$ be the overall interference  and let $\widetilde{P}=P_{2}+\dotsc+P_{K}$ be 
its (per-symbol) average power (see~\eqref{eqn:int_power}). Let $\rvC(\barP )$ be the ``interference-as-noise''
rate where $\barP$ is the SINR of the channel  (see  \eqref{eqn:sinr}). 

We start with the direct part.
Following the steps leading to \eqref{eqn:upper_bd_pe}, we obtain
\begin{align}
&\barp_{\rme, n}\nn\\*
  &\le\Pr\biggl[ \frac{\| \widetilde{X}^n+Z^n\|^2}{2(\tilP+1)} - \frac{\| X_1^n+\widetilde{X}^n+Z^n\|^2}{2(P_1+\tilP+1)} \nn\\*
  & \quad \hspace*{1.25cm} \ge n\rvC(\barP) - \log M -\log(K_0\sqrt{n})\bigg]+\frac{1}{\sqrt{n}}\\
 &\le\Pr\biggl[ ( P_1\!+\!\tilP\!+\! 1 ) \|\widetilde{X}^n +Z^n\|^2 - (\tilP+1) \| X_1^n + \widetilde{X}^n \!+\!  Z^n \|^2 \nn\\*
 & \quad\ge 2( \tilP\!+\! 1)( P_1\!+\!\tilP\!+\! 1 ) \Big( n\rvC(\barP) - \log M -\log(K_0\sqrt{n}) \Big)   \bigg]\nn\\*
 &\qquad \hspace*{5.5cm} +\frac{1}{\sqrt{n}}. \label{eq:ShellK_2}
\end{align}
 As in Section~\ref{sec:direct}, the crucial point is the characterization of the asymptotic distribution of
\begin{align}
 & ( P_{1}+\widetilde{P}+1)\|\widetilde{X}^n+Z^n\|^{2}-( \widetilde{P}+1)\|X^n_{1}+\widetilde{X}^n+Z^n\|^{2}\\
 & =\! P_{1}\|\widetilde{X}^n+Z^n\|^{2}-n( \widetilde{P}+1)P_{1}-2( \widetilde{P}+1)\langle X^n_{1},\widetilde{X}^n+Z^n\rangle \label{eqn:expand_norm}\\
 & =\! P_{1}\bigg(\!n\sum_{i=2}^{K}P_{i}\!+\!\|Z^n\|^{2}\!+\! 2\sum_{i=2}^{K}\langle X^n_{i},Z^n\rangle\!+\! 2\sum_{2\le i<j\le K}\!\!\langle X^n_{i},X^n_{j}\rangle\!\!\bigg) \nn\\*
 &\qquad  -n( \widetilde{P}+1)P_{1}-2(\widetilde{P}+1)\bigg(\langle X^n_{1},Z^n\rangle+\sum_{i=2}^{K}\langle X^n_{1},X^n_{i}\rangle\bigg)\\
 & =\! P_{1}\bigg( \|Z^n\|^{2}-n+2\sum_{i=2}^{K}\langle X^n_{i},Z^n\rangle+2\sum_{2\le i<j\le K}\langle X^n_{i},X^n_{j}\rangle\bigg)\nn\\*
 &\qquad-2( \widetilde{P}+1)\bigg(\langle X^n_{1},Z^n\rangle+\sum_{i=2}^{K}\langle X^n_{1},X^n_{i}\rangle\bigg).\label{eq:long_and_painful}
\end{align}
where \eqref{eqn:expand_norm} follows by expanding the term $\|X^n_{1}+\widetilde{X}^n+Z^n\|^{2}$ and noting that $\|{X}^n_{i}\|^{2}=nP_{i}$ almost surely, and we have used the definitions of $\widetilde{X }^n$ and $\widetilde{P}$.

To apply the Berry-Esseen central limit theorem for functions of random vectors  in Theorem \ref{thm:berry} in Appendix \ref{app:be_func}, we write~\eqref{eq:long_and_painful}
in terms of the zero-mean random variables
\begin{align}
A_{1,\ell} &: =1-Z_{\ell}^{2} \label{eqn:defA1}\\
A_{2,\ell} &: =\sqrt{P_{1}}X_{1,\ell}^{\prime}Z_{\ell}\\
A_{3,\ell} & :=(X_{1,\ell}^{\prime})^{2}-1\\
A_{4,\ell}^{(i)} & :=\sqrt{P_{1}P_{i}}X_{1,\ell}^{\prime}X_{i,\ell}^{\prime}\\
A_{5,\ell}^{(i)} & :=\sqrt{P_{i}}X_{i,\ell}^{\prime}Z_{\ell}\\
A_{6,\ell}^{(i)} & :=(X_{i,\ell}^{\prime})^{2}-1\\
A_{7,\ell}^{(i,j)} & :=\sqrt{P_{i}P_{j}}X_{i,\ell}^{\prime}X_{j,\ell}^{\prime}.\label{eqn:defA7}
\end{align}
Here,  $\{X_{i,\ell}^{\prime}\}$ ,where $\ell \in [1: n]$ and $i\in [2:  K]$, are \iid $\calN(0,1)$-distributed random variables and $j\in [2:K]$.  
Note that the random variables  in \eqref{eqn:defA1}--\eqref{eqn:defA7} are \iid over $\ell$.
We also define the following function:
\begin{align}
& f\Bigl(a_1,a_2, a_3, \bigl\{a_{4}^{(i)} \bigr\}_{i=2}^K, \bigl\{a_{5}^{(i)} \bigr\}_{i=2}^K, \bigl\{a_{6}^{(i)} \bigr\}_{i=2}^K, \bigl\{a_{7}^{(i,j)} \bigr\}_{i,j=2}^K \Bigr)\nn\\*
&:= P_{1}\Bigg(a_{1}-2\sum_{i=2}^{K}\frac{a_{5}^{(i)}}{\sqrt{1+a_{6}^{(i)}}} \nn\\*
&\quad -2\sum_{2\le i<j\le K}\frac{a_{7}^{(i,j)}}{\sqrt{1+a_{6}^{(i)}}\sqrt{1+a_{6}^{(j)}}}\Bigg) \nn\\*
&\quad+2(1+\widetilde{P})\Bigg(\frac{a_{2}}{\sqrt{1+a_{3}}}+\sum_{i=2}^{K}\frac{a_{4}^{(i)}}{\sqrt{1+a_{3}}\sqrt{1+a_{6}^{(j)}}}\Bigg).
\end{align}
With these definitions, we find that 
\begin{align}
&  -n f\Biggl(
  \frac{1}{n}\sum_{\ell=1}^n A_{1,\ell},\frac{1}{n}\sum_{\ell=1}^n A_{2,\ell},
  \frac{1}{n}\sum_{\ell=1}^n A_{3,\ell}, \nn\\*
& \qquad\quad\,\,  \lefto\{\frac{1}{n}\sum_{\ell=1}^n A_{4,\ell}^{(i)}\right\}_{i=2}^{K}, 
  \lefto\{\frac{1}{n}\sum_{\ell=1}^n A_{5,\ell}^{(i)}\right\}_{i=2}^{K}, \nn\\* 
  &\qquad\quad\,\,  \lefto\{\frac{1}{n}\sum_{\ell=1}^n A_{6,\ell}^{(i)}\right\}_{i=2}^{K},
  \lefto\{\frac{1}{n}\sum_{\ell=1}^n A_{7,\ell}^{(i,j)}\right\}_{i,j=2}^{K}
  \Biggr)
\end{align}
coincides with the right-hand side of~\eqref{eq:long_and_painful}

The variances of each of the random variables  in \eqref{eqn:defA1}--\eqref{eqn:defA7}, and the entries of the Jacobian of $f$ with respect to each of its entries, are as follows:
\begin{alignat}{3}
\text{R.V.} &\qquad\text{Variance} &&\qquad\text{Jacobian entry} \notag \\ 
(A_{1,\ell}) & \qquad\xi-1 && \qquad P_{1} \label{eq:first-row}\\
(A_{2,\ell}) & \qquad P_{1} && \qquad 2( \widetilde{P}+1)\\
(A_{3,\ell}) & \qquad 2 &&\qquad 0\\
(A_{4,\ell}^{(i)}) & \qquad P_{1}P_{i} && \qquad 2( \widetilde{P}+1)\\
(A_{5,\ell}^{(i)}) & \qquad P_{i} && \qquad -2P_{1}\\
(A_{6,\ell}^{(i)}) & \qquad 2 && \qquad 0\\
(A_{7,\ell}^{(i,j)}) & \qquad P_{i}P_{j} &&\qquad -2P_{1}.\label{eq:last-row}
\end{alignat}
The covariance matrix $\bV$ of the random vector  $(A_{1,1},A_{2,1}, A_{3,1}, \{A_{4,1}^{(i)} \}_{i=2}^K, \{A_{5,1}^{(i)} \}_{i=2}^K, \{A_{6,1}^{(i)} \}_{i=2}^K, \{A_{7,1}^{(i,j)} \}_{i,j=2}^K)$ is diagonal, and hence we can compute $\mathbf{J}\mathbf{V}\mathbf{J}^{T}$ by
multiplying each of the variances in~\eqref{eq:first-row}--\eqref{eq:last-row} with the corresponding Jacobian entry \red{squared}, and by summing up all resulting terms.
This gives
%
\begin{align}
&\mathbf{J}\mathbf{V}\mathbf{J}^{T}\nn\\*
 & =P_{1}^{2}(\xi-1)+4P_{1}( \widetilde{P} + 1)^{2}+4P_{1}( \widetilde{P} + 1)^{2}\sum_{i=2}^{K}P_{i} \nn\\*
 &\quad +4P_{1}^{2}\sum_{i=2}^{K}P_{i}+4P_{1}^{2}\sum_{2\le i<j\le k}P_{i}P_{j}\\
 & =P_{1}^{2}(\xi-1)+4P_{1}( \widetilde{P} + 1)^{2}+4P_{1}( \widetilde{P} + 1)^{2}\widetilde{P}\nn\\*
 &\quad+4P_{1}^{2}\widetilde{P}+4P_{1}^{2}\sum_{2\le i<j\le k}P_{i}P_{j}\\
 & =P_{1}^{2}\big((\xi-1)+4\widetilde{P}\big)+4P_{1}( \widetilde{P} + 1)^{3}+4P_{1}^{2}\sum_{2\le i<j\le k}P_{i}P_{j}.\label{eq:variance_not_normalized}
\end{align}
To ensure that the effective noise power is one (just as in the case of \iid interference and \iid noise in Section~\ref{sec:prf_inter_iid}), we  divide~\eqref{eq:variance_not_normalized} by $\big(2( \widetilde{P}+1)( P_{1}+\widetilde{P}+1)\big)^{2}$
in accordance with the right hand side of the inequality in the probability in \eqref{eq:ShellK_2}.
This gives the desired achievability part of~\eqref{eqn:interf_shell}--\eqref{eqn:shell_disp2}, and ensemble tightness follows from steps similar to those detailed in Section \ref{sec:prf-she}.

\section{Summary and Future Work} \label{sec:conclu}
In this paper,  \red{we have derived the second-order asymptotic expansion for information transmission over an additive i.i.d. non-Gaussian noise channel when the codebooks are constrained to be  Gaussian} and the decoding rule is constrained to be the NN (or minimum Euclidean distance) rule. We have shown that, under mild conditions on the noise, the dispersion depends on the \iid noise only through its \blue{second and fourth moments}. 
Motivated by the relative simplicity of the implementation of point-to-point codes~\cite{bacelli,Bandemer} we have also leveraged our second-order results to obtain a characterization of the second-order performance of point-to-point codes over Gaussian interference channels.

Some natural questions arise from our analysis and results:
\begin{enumerate}
\item Can one perform similar analyses for the source coding or rate-distortion counterpart~\cite{Lap97}, where source sequences are described by the codeword closest to it in a certain (possibly mismatched) distortion metric? We expect that we need to leverage techniques from finite blocklength rate-distortion theory~\cite{kost12} to obtain  a result similar to Theorem \ref{thm:disp}.
\item Can one extend the analysis in Section \ref{sec:int_network} to the {\em multi-terminal} setting where we are interested in two (or more) rates, say of transmitters $1$ and $2$, \blue{and all the other users' codewords  are treated as noise?}  We envisage  that this analysis would be significantly more challenging due to the lack of a systematic treatment of second-order asymptotics for multi-terminal information theory problems~\cite[Part~III]{TanBook}.
\end{enumerate}
\appendices

\section{Berry-Esseen Theorem for Functions of I.I.D.\ Random Variables}\label{app:be_func}
In this appendix, we state a Berry-Esseen theorem for functions of i.i.d.\ random variables. This theorem is an i.i.d.\ version of a theorem proved by Iri and Kosut~\cite[Prop.~1]{Iri15} for Markov processes.  See  also MolanvianJazi and Laneman's work \cite[Prop.~1]{Mol13c}. 

\begin{theorem}\label{thm:berry}
Let $\bU_i\in\bbR^m$ for $i = 1,\ldots,n$ be zero-mean, i.i.d.\ random  vectors with covariance matrix $\bV = \cov(\bU_i)$ and finite third moment $\bbE[\|\bU_i\|^3]$. Let $f:\bbR^m\to\bbR$ be a function with uniformly bounded second-order partial derivatives in an $m$-dimensional hypercube neighborhood of $\mathbf{0}$. Let $\bJ = (J_1,\ldots, J_m)$ be the vector of first-order partial derivatives at $\mathbf{0}$, i.e., 
\begin{equation}
J_r := \frac{\partial f(\bu) }{ \partial u_r}\Big|_{\bu=\mathbf{0} },\quad r = 1,\ldots, m.
\end{equation}
Let $\sigma^2 := \bJ^T \bV\bJ$ and assume that it is positive.   
Moreover, fix a constant $\alpha>0$. There exists a  constant
$B\in(0,\infty)$ such that for every $\delta$ satisfying $|\delta|\le\alpha$, 
\begin{equation}
\left| \Pr\left( f\Big(\frac{1}{n}\sum_{i=1}^n \bU_i\Big)\ge f(\mathbf{0})+\frac{\sigma}{\sqrt{n}}\delta \right)- \rmQ(\delta)	 \right|\le\frac{B}{\sqrt{n}}
\end{equation}
for all $n \ge 1$. 
\end{theorem}
 
\section{Proof of \eqref{eqn:modZ} }\label{app:moderate}
\blue{By the union bound,} it suffices to show that 
\begin{align}
\Pr\biggl[\frac{1}{n}\|Y^n\|^2 \notin\calP_Y \bigg] &= O \Big( \frac{1}{\sqrt{n}} \Big), \label{eqn:modP1}\\
\Pr\biggl[\frac{1}{n}\|Z^n\|^2 \notin\calP_Z\bigg] &= O \Big( \frac{1}{\sqrt{n}} \Big),\quad\mbox{and}\label{eqn:modP2}\\
\Pr\biggl[\Big(\frac{1}{n}\|Y^n\|^2,\frac{1}{n}\|Z^n\|^2\Big) \notin\calQ\bigg] &= O \Big( \frac{1}{\sqrt{n}} \Big). \label{eqn:modQ}
\end{align}
We first prove~\eqref{eqn:modP1}. 
Note that
\begin{align}
\mathfrak{p}_n & :=\Pr\biggl[ \frac{1}{n}\|Y^n\|^2 \notin\calP_Y \bigg] \\
& =\Pr\biggl[  \bigg|  \frac{1}{n}\|Y^n\|^2 - (1+P) \bigg| > \sqrt{c_y\frac{\log n}{n}}\bigg] \\
& =\Pr\biggl[  \bigg| \frac{2\langle X^n,Z^n\rangle}{n} +  \frac{\|Z^n\|^2}{n} - 1\bigg| > \sqrt{c_y\frac{\log n}{n}}\bigg] \label{eqn:useXas}.
\end{align}
Here, \eqref{eqn:useXas} follows because $Y^n=X^n+Z^n$ and because $\|X^n\|^2 = nP$ almost surely. 
Now recall the definitions of the  random variables $A_{1,i},A_{2,i}$ and $A_{3,i}$ in  \eqref{eqn:defAi1}--\eqref{eqn:defAi3}.  These random variables are \iid across $i \in [1: n]$. Defining
\begin{equation}
g(a_1, a_2, a_3) := -a_1 + \frac{2a_2}{\sqrt{ 1+a_3}},
\end{equation}
one can check that
\begin{multline}
g\lefto( \frac{1}{n}\sum_{i=1}^n A_{1,i}, \frac{1}{n}\sum_{i=1}^n A_{2,i}, \frac{1}{n}\sum_{i=1}^n A_{ 3,i}\right)\\
= \frac{2\langle X^n,Z^n\rangle}{n} +  \frac{\|Z^n\|^2}{n} - 1. \label{eqn:g_clt}
\end{multline}
Furthermore,   the Jacobian matrix of the function $g$ evaluated at $\mathbf{0}$ is $\bJ = [-1\,\,\, 2\,\,\, 0]$ and the covariance matrix $\bV$ of the vector $[A_{1,1}\,\,\, A_{2,1} \,\,\, A_{3,1} ]$ is $\bV=\mathrm{diag}( [ \xi-1 \,\,\, P\,\,\, 2 ]^T )$. 
Thus, by the  Berry-Esseen central limit theorem for functions of random vectors (Theorem \ref{thm:berry} in Appendix \ref{app:be_func}), the random variable ${\sqrt{n} }( 2\langle X^n,Z^n\rangle/n + \|Z^n\|^2/n - 1)$ converges in distribution to a univariate zero-mean  normal random variable with variance $\bJ\bV\bJ^T = \xi-1+4P=c_y$. 
Thus,
\begin{align}
\mathfrak{p}_n  &= \Pr\biggl[ \bigg| \frac{1}{n }( 2\langle X^n,Z^n\rangle + \|Z^n\|^2 - n) \bigg|>\sqrt{c_y \frac{ \log n}{n}}\bigg] \\
&\le 2 \left( \rmQ\bigl( \sqrt{\log n} \bigr) + O\Bigl( \frac{1}{\sqrt{n}}\Big)\right)\label{eqn:be4funcs}\\
&\le O\Bigl( \frac{1}{\sqrt{n}}\Big)\label{eqn:be4funcs_2}.
\end{align}
Here, \eqref{eqn:be4funcs} follows from the union-of-events bound and~\cite[Prop.~1]{Iri15} (this requires the finiteness of the sixth moment of $Z$);  
 the final inequality~\eqref{eqn:be4funcs_2} holds because   $\rmQ(t)\le e^{-t^2/2}$  for $t>0$.   This verifies \eqref{eqn:modP1}.

The proof for the bound in~\eqref{eqn:modP2} is simpler. 
The random variable $( \|Z^n\|^2 - n)/{\sqrt{n} }$ converges in distribution to a univariate, zero-mean normal random variable with variance $\xi-1$. 
The standard Berry-Esseen theorem (again with the assumption that $\bbE[Z^6] < \infty$), together with calculations similar to \eqref{eqn:be4funcs} and \eqref{eqn:be4funcs}, immediately give the desired bound.

Finally, to prove \eqref{eqn:modQ}, we define the random variable
\begin{align}
B&:=\hatP_Y+ P-\hatP_Z  .
\end{align}
Performing calculations similar to those that led to \eqref{eqn:useXas}, we rewrite $B$ as follows:
\begin{align}B
&= \frac{1}{n}\|Y^n\|^2 +P -\frac{1}{n}\|Z^n\|^2  \\
&= \frac{1}{n}\|X^n+Z^n\|^2 +P -\frac{1}{n}\|Z^n\|^2  \\
&= \frac{1}{n}\|X^n\|^2 + \frac{1}{n}\| Z^n\|^2 +\frac{2}{n}\langle X^n,Z^n\rangle+P -\frac{1}{n}\|Z^n\|^2  \\
&=  \frac{2}{n}\langle X^n,Z^n\rangle+2P  . \label{eqn:derivateB}
\end{align}
We now use the argument that led to~\eqref{eqn:g_clt}.
Specifically, let
\begin{equation}
h(a_2, a_3) := \frac{2a_2}{\sqrt{1+a_3}}.
\end{equation}
Then  
\begin{equation}
h\bigg( \frac{1}{n}\sum_{i=1}^n A_{2,i} ,\frac{1}{n}\sum_{i=1}^n A_{3,i} \bigg)=\frac{2}{n}\langle X^n,Z^n\rangle,
\end{equation}
where the i.i.d.\ random vectors $[A_{2,i} \,\,\, A_{3,i} ]$ (for $i \in [1:n]$) were defined in  \eqref{eqn:defAi2}--\eqref{eqn:defAi3}.
 By using standard calculations for the Jacobian of $h$ and for the covariance matrix of $[A_{2,1} \,\,\, A_{3,1} ]$, we see that $\sqrt{n} \cdot  (2\langle X^n,Z^n\rangle/n)$ converges in distribution to a zero-mean Gaussian with variance $4P$. Now, the probability in \eqref{eqn:modQ} can be simplified as follows:
 \begin{align}
 \Pr \bigl[  B \le\eta\big] &= \Pr\lefto[ \frac{2\langle X^n,Z^n\rangle}{n}+ 2P \le\eta \right]\\
&= \Pr\lefto[  \frac{2\langle X^n,Z^n\rangle}{\sqrt{n}} \le \sqrt{n}(\eta-2P) \right]\\
&\le\rmQ\lefto( \frac{ \sqrt{n}(2P-\eta )}{\sqrt{4P}} \right) + O\Big( \frac{ 1}{\sqrt{n}}\Big)\label{eqn:ber}\\
&\le \exp\biggl(- n\cdot \frac{(2P-\eta)^2}{  8P  }\bigg) +  O\Big( \frac{ 1}{\sqrt{n}}\Big)\label{eq:chernoff}.
 \end{align}
 Here,  \eqref{eqn:ber} follows from an  application of  the  Berry-Esseen central limit theorem for functions of random vectors in Theorem \ref{thm:berry}  and~\eqref{eq:chernoff} holds because $\rmQ(t)\leq e^{-t^2/2}$ for $t> 0$. Since $\eta\in (0,2P)$, this proves~\eqref{eqn:modQ}.

\bibliographystyle{unsrt}
\bibliography{isitbib}

\begin{thebibliography}{10}

\bibitem{Strassen}
V.~Strassen.
\newblock {Asymptotische Absch\"{a}tzungen in Shannons Informationstheorie}.
\newblock In {\em Trans. Third Prague Conf. Inf. Theory}, pages 689--723,
  Prague, 1962.
\newblock http://www.math.cornell.edu/$\sim$pmlut/strassen.pdf.

\bibitem{Hayashi09}
M.~Hayashi.
\newblock Information spectrum approach to second-order coding rate in channel
  coding.
\newblock {\em IEEE Trans. on Inform. Th.}, 55(11):4947--4966, 2009.

\bibitem{PPV10}
Y.~Polyanskiy, H.~V. Poor, and S.~Verd\'{u}.
\newblock Channel coding rate in the finite blocklength regime.
\newblock {\em IEEE Trans. on Inform. Th.}, 56(5):2307--2359, 2010.

\bibitem{Sha59b}
C.~E. Shannon.
\newblock Probability of error for optimal codes in a {G}aussian channel.
\newblock {\em The Bell System Technical Journal}, 38:611--656, 1959.

\bibitem{Lapidoth}
A.~Lapidoth.
\newblock Nearest neighbor decoding for additive non-{Gaussian} noise channels.
\newblock {\em IEEE Trans. on Inform. Th.}, 42(5):1520--1529, 1996.

\bibitem{bacelli}
F.~Baccelli, A.~{El Gamal}, and D.~N.~C. Tse.
\newblock Interference networks with point-to-point codes.
\newblock {\em IEEE Trans. on Inform. Th.}, 57(5):2582--2596, 2011.

\bibitem{Bandemer}
B.~Bandemer, A.~{El Gamal}, and Y.-H. Kim.
\newblock Optimal achievable rates for interference networks with random codes.
\newblock {\em IEEE Trans. on Inform. Th.}, 61(12):6536--6549, Dec 2015.

\bibitem{cover72}
T.~Cover.
\newblock Broadcast channels.
\newblock {\em IEEE Trans. on Inform. Th.}, 18(1):2--14, 1972.

\bibitem{Han81}
T.~S. Han and K.~Kobayashi.
\newblock {A new achievable rate region for the interference channel}.
\newblock {\em IEEE Trans. on Inform. Th.}, 27(1):49--60, 1981.

\bibitem{Iri15}
N.~Iri and O.~Kosut.
\newblock {Third-order coding rate for universal compression of Markov
  sources}.
\newblock In {\em Proc. of Intl. Symp. on Inform. Th.}, pages 1996--2000, 2015.

\bibitem{cra99}
H.~Cram\'er.
\newblock {\em Mathematical methods of statistics}, volume~9.
\newblock Princeton University Press, 1999.

\bibitem{Mol13c}
E.~{MolavianJazi} and J.~N. Laneman.
\newblock A second-order achievable rate region for {Gaussian} multi-access
  channels via a central limit theorem for functions.
\newblock {\em IEEE Trans. on Inform. Th.}, 61(12):6719--6733, Dec 2015.

\bibitem{Quoc15}
S.-Q. Le, V.~Y.~F. Tan, and M.~Motani.
\newblock A case where interference does not affect the channel dispersion.
\newblock {\em IEEE Trans. on Inform. Th.}, 61(5):2439--2453, 2015.

\bibitem{KosutSankar1}
O.~Kosut and L.~Sankar.
\newblock Asymptotics and non-asymptotics for universal fixed-to-variable
  source coding.
\newblock {\em Submitted to the IEEE Trans. on Inform. Th.}, Dec 2014.
\newblock {\tt arXiv:1412.4444 [cs.IT]}.

\bibitem{TanTom13a}
V.~Y.~F. Tan and M.~Tomamichel.
\newblock The third-order term in the normal approximation for the {AWGN}
  channel.
\newblock {\em IEEE Trans. on Inform. Th.}, 61(5):2430--2438, 2015.

\bibitem{Cov06}
T.~M. Cover and J.~A. Thomas.
\newblock {\em Elements of Information Theory}.
\newblock Wiley-Interscience, 2nd edition, 2006.

\bibitem{Rice}
S.~O. Rice.
\newblock Communication in the presence of noise--probability of error for two
  encoding schemes.
\newblock {\em The Bell System Technical Journal}, 29:60, 1950.

\bibitem{Stam}
A.~J. Stam.
\newblock Limit theorems for uniform distributions on spheres in
  high-dimensional {E}uclidean spaces.
\newblock {\em Journal of Applied Probability}, 19(1):221--228, 1982.

\bibitem{Tanizaki}
H.~Tanizaki.
\newblock {\em Computational Methods in Statistics and Econometrics}.
\newblock {CRC Press}, 2004.

\bibitem{BahRR60}
R.~R. Bahadur and R.~Ranga Rao.
\newblock On deviations of the sample mean.
\newblock {\em Ann. Math. Statist.}, 31(4):1015--1027, 1960.

\bibitem{Dembo}
A.~Dembo and O.~Zeitouni.
\newblock {\em Large Deviations Techniques and Applications}.
\newblock Springer, 2nd edition, 1998.

\bibitem{CS93}
N.~R. Chaganty and J.~Sethuraman.
\newblock Strong large deviation and local limit theorems.
\newblock {\em Annals of Probability}, 21(3):1671--1690, 1993.

\bibitem{Lap97}
A.~Lapidoth.
\newblock On the role of mismatch in rate distortion theory.
\newblock {\em IEEE Trans. on Inform. Th.}, 43(1):38--47, Jan 1997.

\bibitem{kost12}
V.~Kostina and S.~Verd\'{u}.
\newblock Fixed-length lossy compression in the finite blocklength regime.
\newblock {\em IEEE Trans. on Inform. Th.}, 58(6):3309--3338, 2012.

\bibitem{TanBook}
V.~Y.~F. Tan.
\newblock Asymptotic estimates in information theory with non-vanishing error
  probabilities.
\newblock {\em {Foundations and Trends$\,$\textregistered $ $ in Communications
  and Information Theory}}, 11(1--2):1--184, Sep 2014.

\end{thebibliography}

\subsection*{Acknowledgments}
The authors would like to thank the anonymous reviewers for their careful reading and \blue{helpful} suggestions. The authors would also like to thank  Albert Guill\'en i F\`abregas, Alfonso Martinez, Yury Polyanskiy, and Wei Yang for useful discussions, and Oliver Kosut for bringing our attention to Proposition~1 in~\cite{Iri15}. Finally, VT and GD would like to thank Morten Fjeld and the Chalmers Sweden-NUS Singapore Joint Strategic Project for Education and Research in Human-Computer Interaction (HCI)  for providing support for our research collaboration.

\begin{IEEEbiographynophoto}{Jonathan Scarlett}
(S'14 -- M'15) received 
the B.Eng. degree in electrical engineering and the B.Sci. degree in 
computer science from the University of Melbourne, Australia. In 2011, 
he was a research assistant at the Department of Electrical \& Electronic 
Engineering, University of Melbourne.  From October 2011 to August 2014,  
he was a Ph.D. studen t in the Signal Processing and Communications Group
at the University of Cambridge, United Kingdom. He
is now a post-doctoral researcher with the Laboratory for Information
and Inference Systems at the \'Ecole Polytechnique F\'ed\'erale de Lausanne,
Switzerland.  His research interests are in the areas of information theory, 
signal processing, machine learning, and high-dimensional statistics. 
He received the Cambridge Australia Poynton International Scholarship, and the EPFL Fellows postdoctoral fellowship co-funded by Marie Sk{\l}odowska Curie.
\end{IEEEbiographynophoto}

\begin{IEEEbiographynophoto}{Vincent Y.\ F.\ Tan} (S'07-M'11-SM'15) was born in Singapore in 1981. He is currently an Assistant Professor in the Department of Electrical and Computer Engineering (ECE) and the Department of Mathematics at the National University of Singapore (NUS). He received the B.A.\ and M.Eng.\ degrees in Electrical and Information Sciences from Cambridge University in 2005 and the Ph.D.\ degree in Electrical Engineering and Computer Science (EECS) from the Massachusetts Institute of Technology in 2011. He was a postdoctoral researcher in the Department of ECE at the University of Wisconsin-Madison and a research scientist at the Institute for Infocomm (I$^2$R) Research, A*STAR, Singapore. His research interests include information theory, machine learning and statistical signal processing.

Dr.\ Tan received the MIT EECS Jin-Au Kong outstanding doctoral thesis prize in 2011, the NUS Young Investigator Award in 2014, and was placed on the NUS Faculty of Engineering Teaching Award commendation list in 2016. He has authored a research monograph on {\em ``Asymptotic Estimates in Information Theory with Non-Vanishing Error Probabilities''} in the Foundations and Trends in Communications and Information Theory Series (NOW Publishers). He is currently an Associate Editor of the IEEE TRANSACTIONS ON COMMUNICATIONS.
\end{IEEEbiographynophoto}

\begin{IEEEbiographynophoto}{Giuseppe Durisi} (S'02-M'06-SM'12) received the
Laurea (summa cum laude) and Ph.D.~degrees from
the Politecnico di Torino, Italy, in 2001 and 2006,
respectively. From 2006 to 2010, he was a PostDoctoral
Researcher with ETH Zurich, Zurich,
Switzerland. In 2010, he joined the Chalmers
University of Technology, Gothenburg, Sweden,
where he is currently an Associate Professor and the
Co-Director of the Information and Communication
Technology Area of Advance. He is also a Guest
Researcher with Ericsson, Sweden.

His research interests are in the areas of communication and information
theory. He was a recipient of the 2013 IEEE ComSoc Best Young Researcher
Award for the Europe, Middle East, and Africa Region. He is a co-author
of a paper that received the Student Paper Award at the 2012 International
Symposium on Information Theory, and the 2013 IEEE Sweden VTCOM-IT
Joint Chapter Best Student Conference Paper Award. Since 2015, he has been
on the Editorial Board of the IEEE TRANSACTIONS ON COMMUNICATIONS
as an Associate Editor. From 2011 to 2014, he served as a Publications Editor
of the IEEE TRANSACTIONS ON INFORMATION THEORY. \end{IEEEbiographynophoto}
\end{document}